# Drying behavior of dense refractory castables. Part 2 – Drying agents and design of heating schedules

A. P. Luz[(1),*], M. H. Moreira[(1)], R. Salomão[(2)], M. A. L. Braulio[(3)], V. C. Pandolfelli[(1)]

[(1)] Federal University of São Carlos, Graduate Program in Materials Science and Engineering (PPGCEM), Rod. Washington Luiz, km 235, São Carlos, SP, 13565-905, Brazil.

[(2)] Materials Engineering Department, São Carlos School of Engineering, University of São Paulo - Avenida Trabalhador São-Carlense 400, São Carlos, SP – Brazil

[(3)] 4Cast, Technical Assistance and Consultancy on Refractories, Rua Aristides de Santi, 6, un. 48, São Carlos, SP, 13571-150, Brazil.

*Corresponding author at: t*el.:* +55-16-3351-8601
E-mail: analuz@ufscar.br or anapaula.light@gmail.com

## Abstract

Drying is the most critical process of the first heating cycle of monolithic dense refractories, as the reduced permeability of the resulting microstructure may lead to explosive spalling and mechanical damage associated with dewatering. The first part of this review series pointed out the various drying stages, the role of the binder components and the techniques that can be used to follow the water release in as-cast refractory materials, when they are exposed to heat. Although defining a suitable heating schedule is a great challenge, some tools can be applied to minimize the spalling risks associated with steam pressurization. In this context, this second review article points out *(i)* the main drying agents and how they affect the resulting castables' microstructure (organic fibers, metallic powders, permeability enhancing active compounds, silica-based additives and chelating agents), and *(ii)* the effects related to the procedures commonly applied during the designing of heating routine (i.e., the role of the heating rate, ramp versus holding time), as well as the influence of the castable's dimension on the overall drying behavior. Considering the recent advances regarding the design of refractory formulations and their processing, one may expect that incorporating suitable drying additives into the prepared composition should lead to a suitable and safer water release in such dense consolidated structures. Besides that, novel engineering opportunities, such as the use of in-situ based experimental techniques (i.e., neutron and X-ray tomography) to obtain more accurate data and the development of numerical models, might help in simulating and predicting the steam pressure developed in refractory systems during their first heating. Consequently, instead of designing conservative drying schedules based on empirical knowledge, the novel optimized heating procedures should be based on technical and scientific information.

## 1. Introduction

Refractories are widely applied as lining materials in equipment used for manufacturing metals, glass, cement, and other products, and that operates at high temperatures [1]. To keep their competitiveness and production rate, these industrial sectors work on a daily basis with a strong pressure to get units back into service and, often, not the required attention is given to the equipment´s heat up procedures. Nevertheless, using proper heating schedules is an important aspect to: *(i)* prevent likely damage to the refractories that can take place even before reaching the operating conditions; and *(ii)* inhibit premature deterioration of ceramic lining systems, leading to unexpected time-consuming repairs required before or during scheduled overhauls.

The refractories' drying process is based on the heat application to the installed lining to withdraw water from the consolidated structure without causing damage to it. Part 1 of this review [2] addressed the various stages in detail, the role of the binder components and the techniques that can be applied to follow the water flow when heating such ceramic materials. As pointed out by some authors [3–9], when refractories are subjected to a first thermal treatment, where it may be the initial heat up after a new lining installation or a subsequent treatment of an existing one after a halt, the drying schedules (ramp rates and holding times) should always be thoroughly discussed and agreed upon the end-user, lining designer, lining installer, refractory manufacturer and dry-out subcontractor.

Selecting appropriate heating schedules is not an easy task and, most of the time, they are usually based on empirical knowledge. Hence, it is very common to apply conservative heating routines during the first castable's thermal treatment, which is an expensive, time-consuming and undesirable procedure as it delays resuming the operation of industrial equipment. Aiming to minimize the impact of an inappropriate drying step, by ensuring safe and faster water release, some routes can be suggested to adjust the castable's microstructure [5,10–18]: *(i)* adding drying agents (i.e. organic fibers, metallic powders, and others) to the ceramic compositions to induce the generation of permeable paths to expel water towards the lining's surface; *(ii)* replacement of traditional hydraulic binders by colloidal ones, inhibiting the hydrate formation and also increasing the refractory permeability [19–21], *(iii)* incorporating metallic fibers into the refractories to improve their mechanical strength and fracture energy, allowing them to withstand higher thermomechanical stresses without cracking, and others [14,22].

Considering these aspects, this article is the second part of a review paper that points out the main drying steps and provides guidance to deal with them without damaging the castable's structural integrity. In the present paper, the following subjects will be discussed in the forthcoming sections: *(i)* main drying agents and how they affect the resulting refractories' microstructure (organic fibers, metallic powders, permeability enhancing active compounds, silica-based additives and chelating agents), and *(ii)* design of drying schedules for different refractory compositions (i.e., the role of the heating rate, ramp versus holding time routine, and the castable's dimension on the overall drying behavior).

**2. Main drying agents and how they influence the castables' properties**

Various routes can be applied to minimize the impact of an inappropriate drying step for the refractory lining. For instance, one may optimize the selected dry out schedule, by inducing higher water withdrawal during the evaporation period (below 110°C), preventing the pressurization of castables at temperatures higher than 110°C [2–4,15,23–28]. To achieve such a performance, some microstructural engineering tools must be considered to speed up the water release (i.e., by designing products with permeable structures) or increase the refractories' mechanical strength.

In general, decreasing the refractories' density (by using porous raw materials) results in an increased water demand during their processing (Table 1) and, therefore, typically higher permeability can be achieved. To consider the extremes, a completely permeable lining will prevent steam pressure buildup and no dry-out concerns should be raised, as the spalling would not take place without the structure's pressurization. Conversely, an impermeable material will not allow any steam to be conveyed to the lining external surface and the integrity of the refractory will depend mainly on its ability to contain the buildup of steam pressure within [6,29–31]. In practice, even residual water confined in the structure can develop pressures that exceed the mechanical strength of the refractory materials, leading to significant damage from explosive spalling.

Table 1: Characteristics of different refractory castables [4,6].

| Properties | Type of refractory castable | | | | |
| --- | --- | --- | --- | --- | --- |
| | Low Cement (CaO = 1.0-2.5 wt.%) | Dense conventional (CaO > 2.5 wt.%) | Medium weight | Insulating | Ultra-light insulating |
| **Density (g/cm$^3$)** | 2.24-2.88 | 2.08-2.72 | 1.60-2.08 | 0.48-1.60 | 0.32-0.48 |
| **Water content (wt.%)** | 4-8 | 8-12 | 15-18 | 20-40 | 40+ |
| **Permeability (after air cure)** | Low | → | | | High |
| **Water release issues** | More difficult | → | | | Less difficult |

Therefore, aiming to adjust the permeability of well-packed castable structures, the following approaches can be taken into consideration: *(i)* addition of organic fibers or aluminum powder to the refractories, which generates permeable paths to expel water vapor towards the ceramic lining's surface [15,23,32,33]; *(ii)* use of additives that are able to interact with the binder compound contained in the castables' formulation, changing the likely hydration reactions sequence and modifying the overall permeability of the molded pieces [34–37]; and *(iii)* the selection of compounds that may inhibit the hydrates formation, preventing the decrease of the material's porosity and permeability [38–40].

The following sections will discuss these approaches and present in more details the mechanisms associated with each additive used to enhance the drying ability of dense refractory castables.

*2.1. Polymeric fibers*

Fibers have been used in refractory products for various purposes (Table 2), mainly to obtain insulating properties or to improve the permeability and reinforce dense monolithics [11,15,41–44]. Different types of fibers are currently available on the market, but the organic ones are the most commonly used in castables when thinking about improving their drying performance. Nevertheless, the selection of the type, shape and amount of these additives requires caution, as these aspects must be related to the particle size distribution and processing parameters of the designed refractories [7,45].

Table 2: Objectives and effects of adding fibers to refractories (adapted from [11]).

| Purpose | Expected effects of the fiber | Type of fiber |
|---|---|---|
| Insulation | Bulk density reduction<br>Porosity increase<br>Formation of insulating layer | Organic fiber<br>Rock wool<br>Ceramic fiber |
| Flexibility | Tensile strength increase<br>(pull-out and bridging) | Metallic |
| Fracture resistance<br>Reliability | Tensile strength increase<br>Fracture energy increase<br>Fracture toughness increase<br>(pull-out and bridging) | Oxide ceramic<br>Metallic<br>Carbon<br>SiC |
| Gas permeability | Controlled permeability increase | Organic |

In general, organic fibers are added to the dry raw materials of the refractory's formulation and, after the casting and setting of the molded pieces, the space occupied by such polymers is preserved, resulting in a large number of fibers scattered per volumetric unity of castable [46]. During the first heat-up, these additives experience several transformations (dissolution, shrinkage, melting and/or burnout), resulting in a high number of permeable paths that connect themselves with matrix-aggregate interfaces, closed pores, as well as dense zones of the matrix region with the surface of the solid structure. Consequently, such channels allow the release of the pressurized water vapor and reduce the explosive spalling likelihood of such ceramic materials.

Among the permeability-increase mechanisms associated with polymeric fibers, three of them may be pointed out as follows:

(i) *Dissolution in water* – for instance, poly(vinyl alcohol) (PVAL) is a hot-water soluble material and, although its fibers are inert during the castables' mixing step, they can dissolve during the initial heating stages when in contact with the hot water contained in the consolidated structure. Hence, the resulting polymeric solution is absorbed by the surrounding matrix leaving partially empty paths [46,47].

(ii) *Swelling when in contact with water* – vegetal fibers (i.e., cellulose, jute) are comprised by tubular structure with empty channels after their drying (Fig. 1a). When they are added to an aqueous environment, such as the one obtained during the castables' mixing and curing steps, strong capillary forces will induce the water to return to the fiber structure, which restores their original larger cross section (Fig. 1b). During the refractory's curing stage, the volume of swollen fibers is preserved, but it will be decreased and helps to generate permeable paths when the consolidated body is dried below 110°C (Fig. 1c) due to water evaporation. After that, the fibers' burnout should take place above 220°C (Fig. 1d), which leads to another significant increase in the microstructure's permeability [5,48].

(iii) *Melting and burnout* – this is the most common mechanism, which is based on the melting and burnout of thermoplastic polymers (i.e., polypropylene, polyethylene, polyolefin-based copolymers, etc. [12,13,15,49–51]), as illustrated in Fig. 2. These fibers are inert during the mixing and curing stages (Fig. 2a) of the castables, but the polymer molecules will become soft and deformable by the pressurized water vapor during their melting (Fig. 2b), allowing the formed gas to flow through the permeable paths partially opened in the microstructure. Considering the temperature increase, the fibers' thermal decomposition should take place, generating fully accessible permeable paths (Fig. 2c) [46].

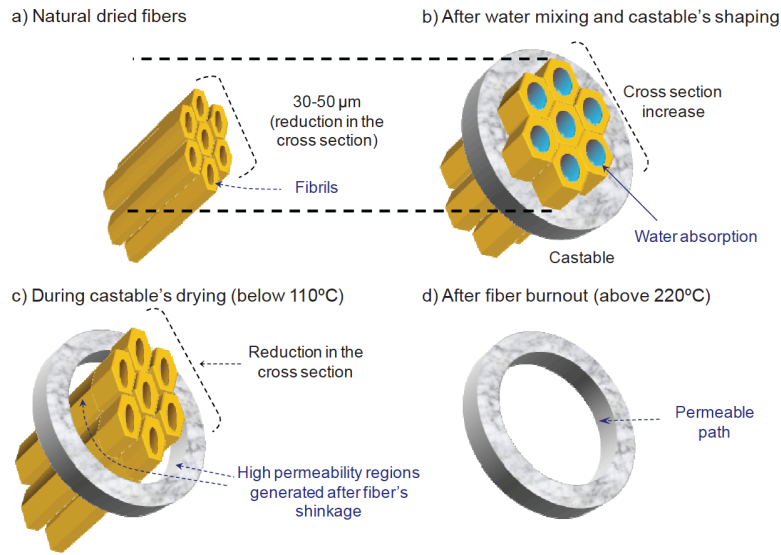

Figure 1: Mechanism of permeability increase for vegetal fibers: (a-c) at room temperature due to the contact with water and (d) after fiber burnout (adapted from [48]).

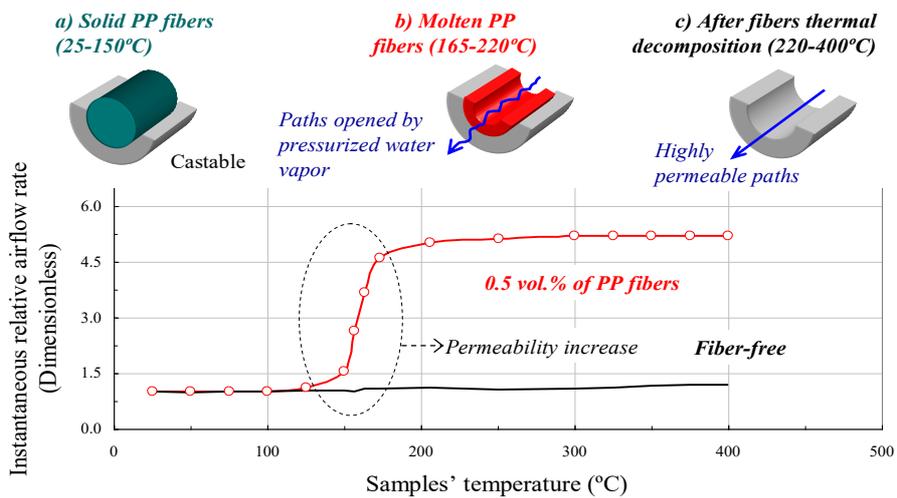

Figure 2: Fluid dynamic behavior of polypropylene (PP) fibers containing refractory castables (hot air permeametry) and schematic representation of fibers' permeability increase mechanism (a-c) [46].

Besides the chemical type (composition), the geometry (shape and size) and the content of the fibers need to be properly selected to optimize the castable's permeability and speed up water withdrawal.

### 2.1.1. Types of fibers added to refractory castables to enhance their drying behavior

The key aspect to be considered when selecting polymeric fibers is that their decomposition should take place at temperatures below those at which pressure build up is observed in the castable's structure. Hence, the effectiveness of the fibers as drying agents depends on their melting and degradation behavior under heating [11,14,18,42,52].

Among the most evaluated types of polymers used for this purpose, the following can be pointed out:

- Polypropylene (PP) and polyethylene terephthalate (PET), which are synthetic materials widely applied in the textile industry. The low cost, high stiffness and tensile strength of PP fibers have made them one of the preferred materials for concrete reinforcement [14,22,51,53]. Moreover, these polymers have the ability to shrink and melt in the 120-170°C (PP) and 200-270°C (PET) temperature ranges [15,54,55].

- Cellulose (polyanhydroglucose) is an important material that occurs naturally in a fibrous state (i.e., cotton, flax, etc.). As an example, dry jute fibers may comprise 12-14 wt.% of lignin, 58-63 wt.% of cellulose and 21-24 wt.% of hemicellulose. After interaction with water, they can swell up to 22%, which make them a good option to act as permeability-aiding additive in refractories. Unlike PP or PET, jute and cellulose fibers do not melt, but rather char above 200°C and 300°C, respectively. Besides that, they present suitable dispersion in aqueous medium, minimizing the likelihood of affecting the overall rheology behavior of the castables [45,48].

- Meta-aramid fibers (PAr) represent a class of high-performance synthetic organic polymers with enhanced flame resistance, used as fabrics and felts in filtration and insulation applications. Because their carbonization process only starts at temperatures above 370°C, they can act inducing a mechanical reinforcement effect at the early heating up stages of the refractories and only increase the overall permeability at higher temperatures [18,42].
- Olephynic copolymer (OC) and linear low-density polyolephynes (LLDPO) are materials with a very low melting point (around 75 and 110°C, respectively), which makes them suitable options to be applied as drying agents to optimize the permeability of dense castables [56].
- Recycled polymers (polypropylene and polyethylene) are also good options to be used in refractories, as physico-chemical changes are commonly observed after their second or third melting and shearing cycles, leading to materials with lower levels of rigidity, heat distortion, melting temperatures, etc. [54,55]. Due to these aspects, in most cases, they can be used in different applications. Nevertheless, their low melting point and easy thermal degradation make them suitable to be used as fibers in refractory products [18,42,46].

Most of the published papers presented in the literature explore the use of PP and PET fibers to adjust the drying behavior of refractories [15,20,38,40,46,56–60]. Additionally, aiming to analyze the efficiency of these additives, "macro"-TGA (thermogravimetric analyses) [20,38,40,50,57,61,62] and cold and hot air permeability measurements [13,18,20,34,41,42,49,62–64] are commonly applied for the evaluation of the castables' performance. In the literature, permeability is usually expressed as constants that are derived from two main models: Darcy's law and the Forchheimer equation (Table 3). The basic difference between them is the type of dependence that the fluid pressure assumes when related to its velocity [65,66]. For practical purposes, it is usually defined that the higher the $k_1$ and $k_2$ values, the more permeable the material's structure is. Moreover, measurements of the air flow rate are also carried out by some researchers as this parameter indicates how easily the fluid can interact and be conveyed through the consolidated microstructure [42,46,48].

Table 3: Equations used for permeability evaluation of refractories [67].

| Model | Equations* | |
|---|---|---|
| | **For incompressible fluids** | **For compressible fluids** |
| Darcy | $\dfrac{P_i - P_0}{L} = \dfrac{\mu}{k_1} v_s$ | $\dfrac{P_i^2 - P_0^2}{2PL} = \dfrac{\mu}{k_1} v_s$ |
| Forchheimer | $\dfrac{P_i - P_0}{L} = \dfrac{\mu}{k_1} v_s + \dfrac{\rho}{k_2} v_s^2$ | $\dfrac{P_i^2 - P_0^2}{2PL} = \dfrac{\mu}{k_1} v_s + \dfrac{\rho}{k_2} v_s^2$ |

* where $P_i$ is the fluid pressure at the sample entrance (Pa); $P_0$ is the fluid pressure at the sample exit (Pa); $v_s$ is the fluid velocity, given by the volumetric rate (Q) divided by the specimen cross-sectional area (A), orthogonal to the fluid flow (m/s); L is the sample thickness (m); $\mu$ is the viscosity of the fluid (Pa.s); $\rho$ is the density of the fluid (kg/m³); $k_1$ is the Darcian or viscous permeability constant (m²); $k_2$ is the non-Darcian or inertial permeability constant (m); and P the fluid pressure at which $v_s$, $\mu$ and $\rho$ are measured or calculated (Pa).

Aiming to highlight the effects of adding fibers to the drying behavior of refractory castables, some data will be presented in this section. For instance, Table 4 summarizes the main features observed while analyzing high-alumina castables bonded with calcium aluminate cement (CAC) and containing different amounts and types of fibers. According to these results, polypropylene and jute fiber incorporation into the refractories resulted in an earlier permeability increase, proving to be more effective in reducing the pressurization risk of the samples. The gain in permeability in both systems was quite similar (360-370%) when compared to the reference material (fiber-free one). Although the permeability increase obtained for the PET-containing samples was in the same range as the ones detected for jute and PP, the higher thermal stability of the former only favored a raise of the air flow above 250°C [42]. This aspect is very relevant, because if polymer's melting and decomposition takes place above the critical temperature (when the mechanical strength of the refractory is not enough to withstand the pressurization level), the castable can explode even before the action of the added fibers is identified. Therefore, among the tested fibers, cellulose was the one that did not perform well, as no major effect on permeability was noticed up to its almost complete degradation at 460°C.

Table 4: Main features observed for high-alumina fiber-containing castables by the hot air permeametry (HAP) technique (adapted from [42]).

| Fiber type | Added amount (vol%) | Starting temperature of permeability increase (°C) | Gain in permeability compared to the fiber-free sample (from room temperature up to 700°C, %) |
|---|---|---|---|
| PP | 0.18 | 180-200 | 360 |
| PET | 0.22 | > 400 | 350 |
| Cellulose | 0.22 | > 460 | 210 |
| Jute | 0.22 | 120 | 370 |
| PAr | 0.23 | > 400 | 150 |

When comparing the most promising fibers (PP and dry jute = named in Fig. 3 as natural fiber), it was observed that a permeability increase was already identified for the samples obtained at the green stage and not only after their thermal treatment at high temperature (Fig. 3a). The non-Darcian permeability constant ($k_2$) for the natural fiber (NF)-containing castable was roughly one order of magnitude higher than the fiber-free or PP-containing ones after curing at 8°C for 15 days under relative humidity (RH) = 100%, followed by 96 h at this temperature with RH = 5% [48]. This $k_2$ increase was attributed to the morphological effects of the NF fibers that took place when water was withdrawn from their structure [18]. The faster water release of NF samples could also be observed by TG evaluation (Fig. 3b), which pointed out the technological potential of this natural material to minimize the castables' spalling risk.

Another interesting comparison among different fibers was carried out by Salomão et al. [56]. These authors analyzed the performance of polypropylene (PP), linear low-density polyolephyne (LLDPO) and olephynic copolymer (OC) when added, in an equal volumetric amount (0.36 vol%) and with close dimensional values (length ~3.89-4.58 mm and diameter ~15-24 μm), to high-alumina CAC-bonded castables. The incorporation of LLDPO and OC fibers into the formulations induced an increase in the number of permeable paths (Fig. 4b) at lower temperatures (110 and 80°C, respectively), allowing the water release before the beginning of the water-vapor pressurization (typically > 110°C, Fig. 4a). Consequently, the drying rates obtained for LLDPO and OC samples raised to levels much higher than those observed for PP and fiber-free compositions and the total time of this step was reduced to half of the usual one (which can be inferred by Fig. 4a as the heating rate was the same for all samples and, thus, a lower temperature indicates a shorter time) [48].

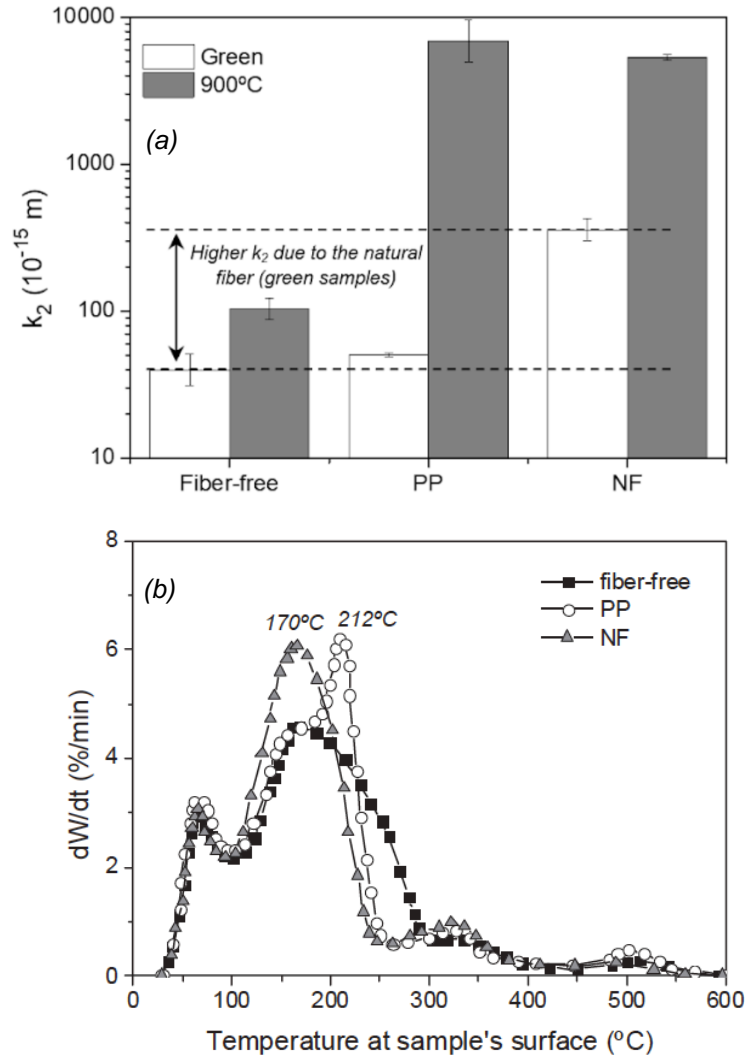

Figure 3: (a) Permeability constant ($k_2$) of green and fired samples (900°C) [18] and (b) drying rate for green and humid castables containing polypropylene (PP) or natural fibers (NF). A heating rate of 10°C/min was applied during the thermogravimetric measurements [48]. Dashed lines presented in (a) points out the permeability increase derived from the NF action when compared to the reference composition.

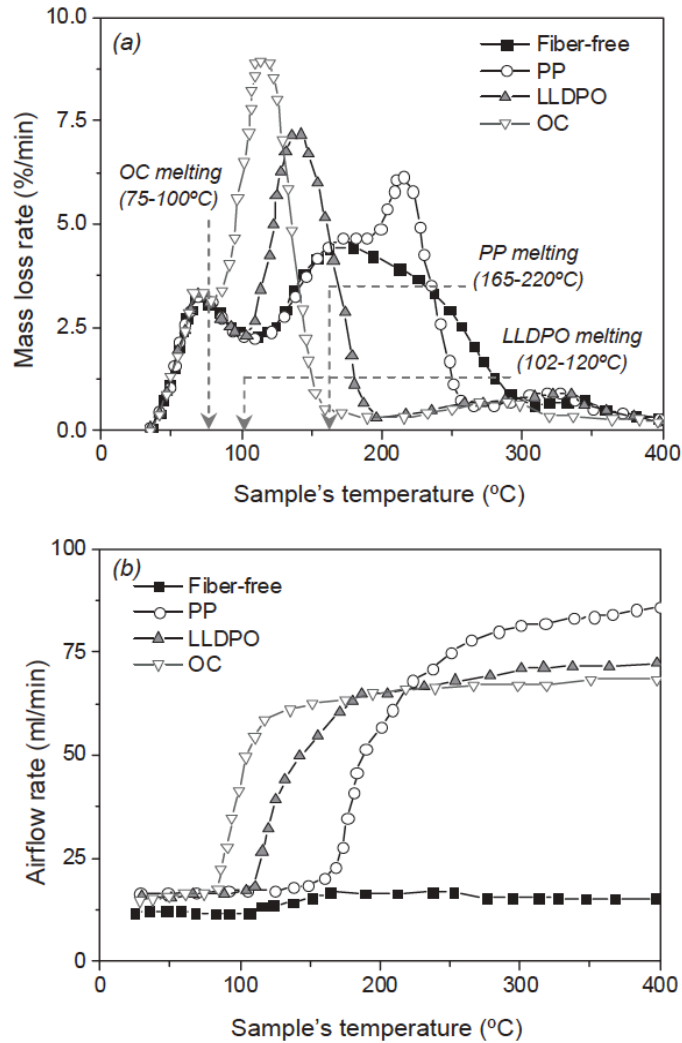

Figure 4: (a) Mass loss rate and (b) airflow rate (obtained via hot air permeametry measurements) of castables containing PP (polypropylene), LLCPO (linear low-density polyolephyne) and OC (olephynic copolymer) fibers [56].

A more recent study [46] also investigated the likelihood of producing fibers from recycled PP and high-density polyethylene (HDPE) drinking straws and applying them as anti-spalling additives in self-flowing high-alumina CAC-bonded castable. A total of 0.36 vol% of fibers (diameter = 23.5-83.4 µm and length = 4.3-5.6 mm) was added to the refractories and the molded samples were cured at 10°C for 15 days under RH = 100%. As indicated in Fig. 5a, fiber-free specimens presented a typical drying behavior, where the water was withdrawn according to the three traditional drying steps (evaporation, ebullition and dehydration). Based on the hot air permeametry measurements (Fig. 5b), adding fibers from the recycled straws (R-Straw-PP and R-Straw-PE) induced the beginning of the samples' permeability increase in the range of 141-161°C. It was stated that, the small but noteworthy differences regarding the maximum airflow and drying rates obtained for the evaluated compositions (containing plain and recycled fibers) were related to the different average diameters and lengths of these additives after the castables' mixing step [46]. Nevertheless, the composition containing R-Straw-PE showed faster permeability and drying rate increase at a much lower temperature (141°C, Fig. 5) than the conventional fibers produced from unprocessed polymers. These results highlight that recycled straws can also be used as feedstock to produce drying additives for refractory castables.

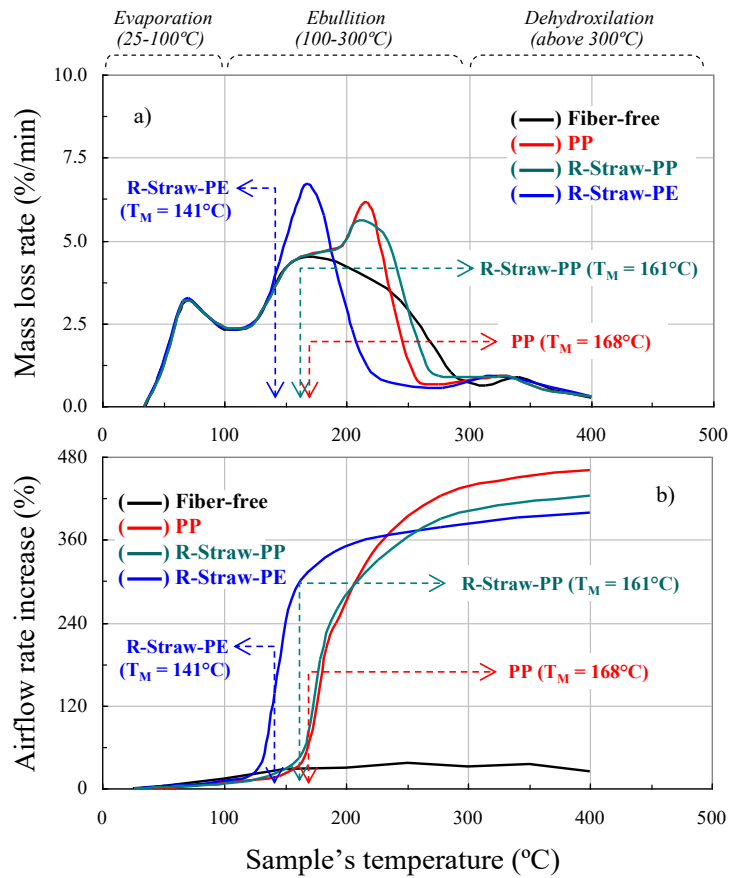

Figure 5: (a) Drying and (b) fluid dynamic behavior of refractory castables containing the fibers produced from unprocessed polypropylene (PP), from recycled PP straws (R-Straw-PP) and recycled HDPE straw (R-Straw-HDPE) (the fiber-free composition is also shown as a reference) [46].

According to the examples presented above, selecting the fiber type is of utmost importance as it will define the temperature range in which permeability will be improved. Additionally, the design of the refractory formulations (which defines the amount of the coarse and fine ceramic particles and how well packed the final microstructure will be) will play a role when selecting these anti-spalling additives. Therefore, the following sections discuss the influence of the particle size distribution (PSD) and geometrical aspects of the fibers in enhancing the drying behavior of dense castables.

### 2.1.2. How the particle size distribution (PSD) of the refractory castables influences the fibers' selection

Refractory producers usually make efforts to design castables that can be placed with a minimum amount of water, as most of the added liquid will be further converted into porosity upon heating. Thus, an important criterion is to adjust the particle size distribution of the compositions to obtain a well packed structure, reducing the need of high-water content for mixing and casting. Nevertheless, this enhanced particle packing is responsible for the spalling of such materials during their subsequent drying step, as it inhibits the steam release from the resulting microstructure.

To develop a specific microstructure, one may be able to play with different raw materials and optimize their contents and grain sizes [68,69]. To do that, particle packing models, such as the ones proposed by Andreasen [70,71] and Dinger and Funk [72,73], are commonly applied for the design of refractory formulations. The main parameters considered in those models are $D$, $D_s$ and $D_L$ (the actual, the smallest and the largest particle size, respectively, among the selected raw materials), as well as $q$, which is the distribution modulus that defines the coarse and fine component (matrix) ratio contained in the composition. Thus, the higher the $q$ value, the greater the aggregate content in the simulated mixtures will be (Fig. 6) [32,70,71].

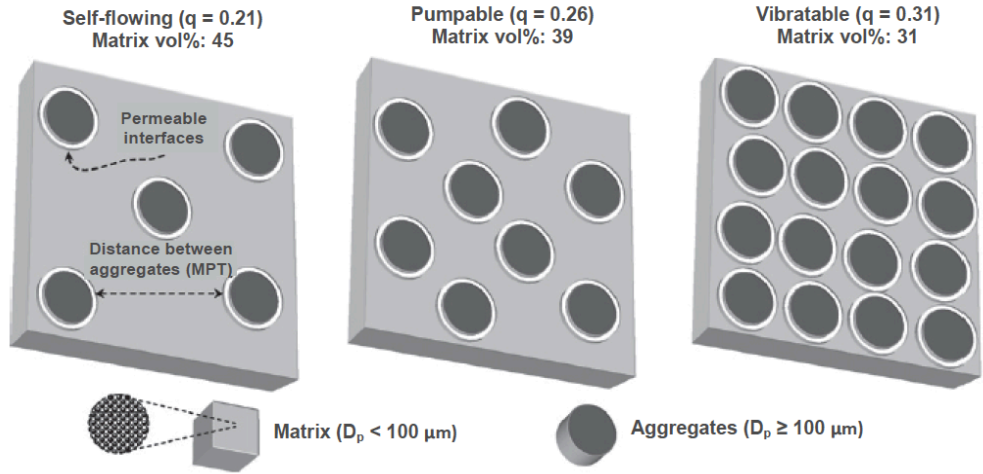

Figure 6: Schematic view of the particle size distribution (PSD) in different classes of refractory castables pointing out the amount of coarse aggregates, the volume content of fine particles (matrix fraction) and permeable interfaces observed in the resulting microstructure [32].

It is accepted that the flow of a gas or liquid through the consolidated refractory structure can commonly be carried out by two main permeable paths. The first one is the matrix-aggregate interfaces, generated by the packing flaws due to the differences of the particle size (also known as *wall effect*) [32,63]. The number of these interfaces can be estimated by the maximum paste thickness (MPT, which represent the distance between aggregates) parameter, as a small value of it indicates that the coarse grains are close to each other (Fig. 6).

The second and most important route for the permeant flow is through the porosity in the castable matrix [67]. As these permeable paths are thin and tortuous, the larger mean distance among fine particles or the interparticle separation distance (IPS) plays a major role in the drying behavior of refractories. It is stated that compositions with higher IPS should present an easier fluid flow throughout the solid structure. For example, self-flowing, pumpable and vibratable castable products present typical matrix/aggregate volumetric ratio between 45-39%, 39-34% and 31-29%, respectively, and the differences in their permeability can be up to two orders of magnitude, as illustrated in Fig. 7 [32,63,74].

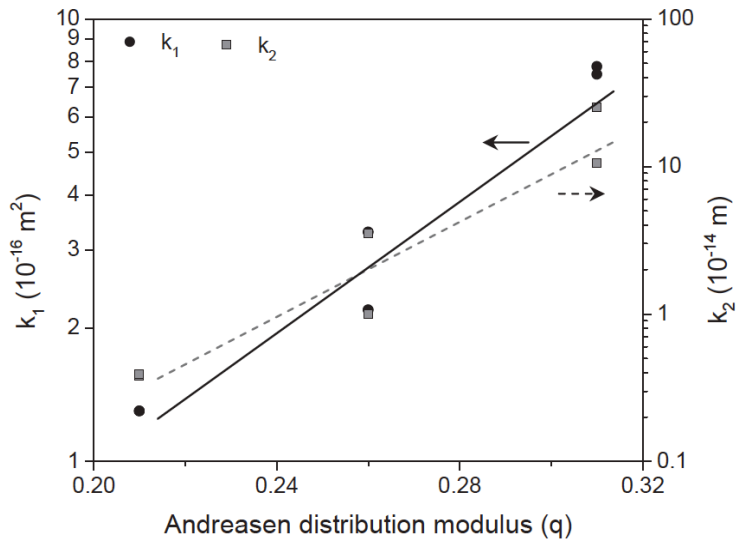

Figure 7: Influence of PSD on the permeability constants of ultra-low-cement (1 wt.%) refractory castables. The samples were prepared with a fixed amount of water = 14 vol% [74].

Besides the IPS and MPT parameters, another important aspect is the packing gradient commonly observed along the castable's thickness. Fig. 8 shows that $k_1$ and $k_2$ values of high alumina-based castables can change from the surface to the bulk of the samples. Dehydration, sintering and microcracking were the key mechanisms causing the variation in the permeability with temperature [74]. Besides that, the castable's upper face (the one exposed to the environment during casting) usually presents a thin layer with distinct particle packing, inducing a better accommodation of fine matrix particles around the aggregate grains and the decrease of the porosity available for fluid flow. This effect led to the lower $k_1$ and $k_2$ values for the samples withdrawn from this upper region (Fig. 8).

From a technological point of view, the existence of a layer of lower permeability at the refractory's surface may negatively affect its drying performance. On the other hand, damage caused to this thin protective layer can also expose the inner structure, which is more susceptible to the penetration of corrosive liquids. Thus, the optimization of the fluid dynamic properties of these products must be mainly based on particle size distribution and/or the addition of polymeric fibers [45].

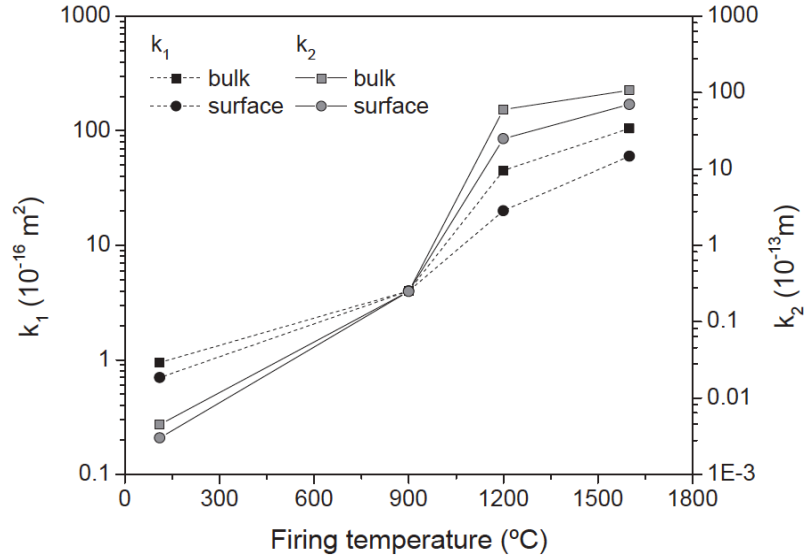

Figure 8: Variation of surface and inner permeability as a function of the firing temperature of a self-flowing high-alumina castable composition (q = 0.21) [74]

In order to increase the number of connections throughout the matrix and among the permeable matrix-aggregate interfaces, it should be considered that polymeric fibers must fulfill two main requirements [32]: *(i)* be added in a proper volumetric amount and *(ii)* present a suitable geometry to interconnect the interparticle zones (ITZs) and the paths generated during their first heating stage. These conditions assure that, for a fixed fiber load, the number of formed permeable interconnections and their performance will be maximized.

When comparing the fiber containing refractories with distinct particle size distributions and placing features (Fig. 9), significant differences in the $k_2$ values could be detected for the green samples, which are only associated with the PSD modifications (*q* changed from 0.21 to 0.31) and the resultant packing (> $k_2$ for > *q*) [32]. The fiber addition did not affect the permeability of the castables under this evaluated condition, as the measurements were carried out after drying the specimens (110°C for 24h) and before the PP fiber decomposition. After firing, on the other hand, the self-flowing composition only presented a permeability increase when the longer fibers (3 mm) were used. The other castables (pumpable and vibratable) showed improvements in their permeability with both fiber lengths. Therefore, the performance of PP fibers depends on a suitable match between fiber length and the PSD.

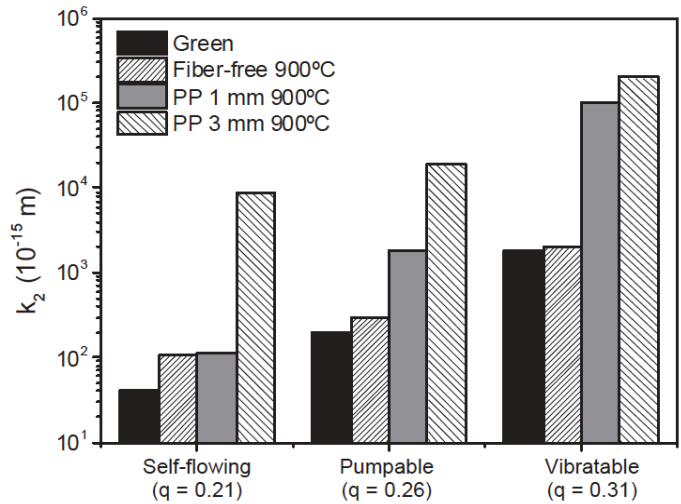

Figure 9: Combined effect of PSD and fiber length on castable's permeability (adapted from [32]). A total of 0.36 vol% of polypropylene fibers (1 mm or 3 mm long and 15 μm diameter) was added to the refractories.

Salomão et al. [32] stated that the maximum paste thickness (MPT) or the average distance between aggregates can be used to estimate the minimum fiber length that will be able to connect the permeable interparticle zones and generate new paths for the steam release. As illustrated in Fig. 10, the addition of 1 mm fibers to a self-flowing high-alumina castable led to no connections and they most likely behaved as closed pores. This might explain why the $k_2$ increase was not identified for the formulation with q = 0.21 containing PP = 1 mm (Fig. 9). This behavior is related to the calculated MPT (for aggregates fraction > 500 μm) of the designed compositions, which resulted in values equivalent to 2.12 mm (q = 0.21), 1.25 mm (q – 0.26) and 0.98 mm (q = 0.31). Thus, the permeability increase only took place when the fiber length was similar or greater than the coarse particle separation distance.

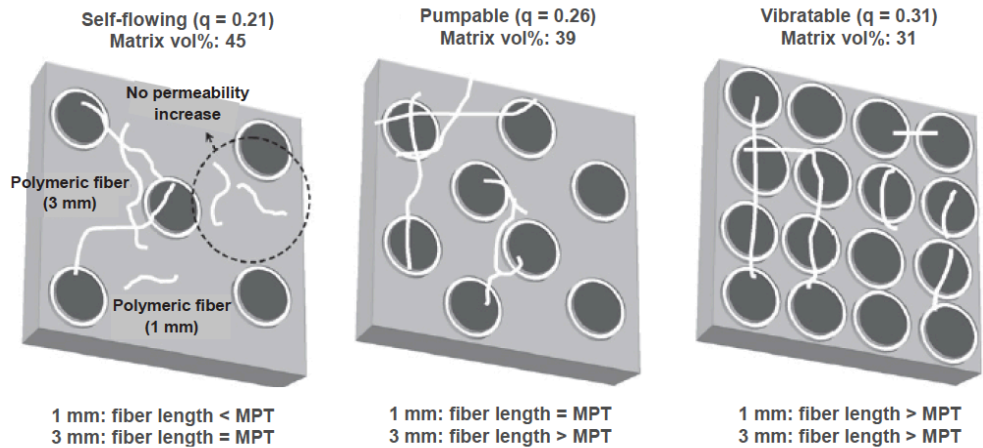

Figure 10: Effect of different length polymeric fiber addition to castables with distinct PSD and their relation with MPT [32].

### 2.1.3. Geometrical aspects of the fibers

Geometrical aspects of the fibers (length and diameter) need to be considered to maximize the connections between low and high permeable regions of the castable's microstructure for the same fiber load added to the formulations. As pointed out in the previous section, for a constant diameter, changes in the additive length modify the number of resultant paths, as a minimum value is required to ensure this positive effect on the overall permeability of the samples.

Fig. 11 shows the $k_1$ values obtained for self-flowing high-alumina castables containing polypropylene fibers of initial length (before mixing) ranging from 0.1 to 24 mm long and 15 μm diameter. According to these data, an increase in the samples' permeability were obtained for fibers longer than 3 mm when compared to the shorter ones (0.1, 0.5 and 1 mm) and after the firing step at 900°C. Nevertheless, adding longer fibers (> 3 mm) did not necessarily enhance the fluid flow behavior as these additives are more likely to be fragmented during the castable's mixing, which generates a higher content of fine fibers after this processing step [75].

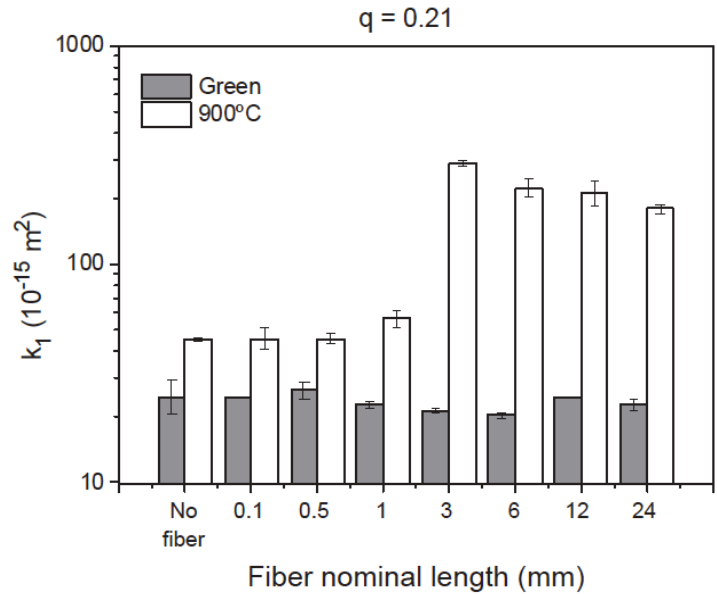

Figure 11: Permeability versus fiber length for self-flowing castables (q = 0.21) [75].

To illustrate this, Fig. 12 presents images of the as-received PP fibers and the ones extracted by flotation from castables (compositions containing 0.09 wt.% of fibers) after their mixing stage. As observed, significant damage to the fiber structure took place due to attrition, shear and stretching forces [76]. Besides that, the higher the fiber length, the greater the dimensional reduction detected after mixing (a decrease of only 10% for 1 mm fiber and 92% for 24 mm one, Table 5). Hence, the changes in the fiber length during use is another aspect that makes selecting these additives more challenging to optimize the number of permeable paths in dense structures.

In addition to the length, changes in the fibers' diameter have also been reported [75]. Although longer fibers decrease the number of paths per volume or mass of castable (Fig. 13a), it ensures the interconnectivity among the permeable regions. Similarly, a reduction in the number of paths per volume was observed for thicker fibers (Fig. 13a and 13b), resulting in a poor distribution of channels throughout the castable's structure, which is less effective in increasing permeability.

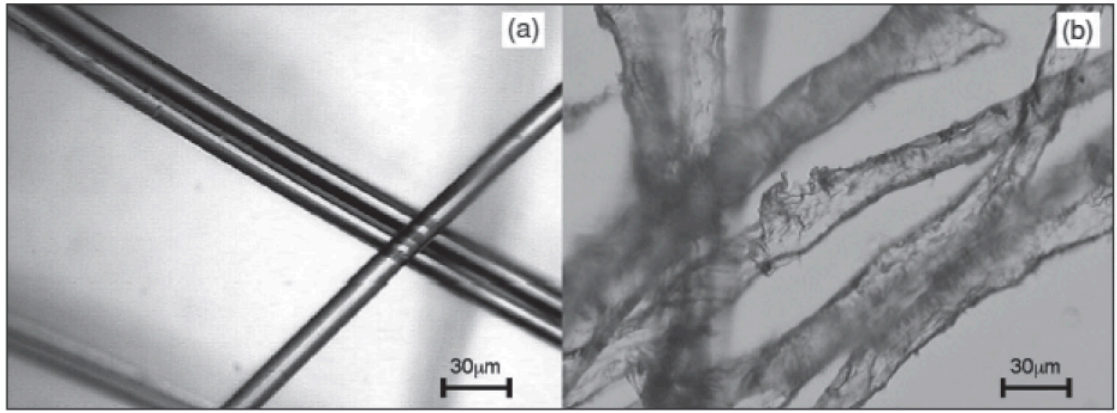

Figure 12: Optical microscopy images of the propylene fibers: (a) as received (smooth, translucent surface) and (b) after mixing step (opaque, scratched surface) of the refractory castables [76].

Table 5: Fiber length reduction observed during the mixing step of refractory castables [76].

| Nominal length (mm) | Average length of as received fiber (mm) | Average length after castable's mixing (mm) | Average length reduction after mixing (%) | Percentage < 1 mm after mixing (%) |
|---|---|---|---|---|
| 0.1 | 0.32 ± 0.25 | 0.51 ± 0.24 | | |
| 0.5 | 0.56 ± 0.21 | 0.61 ± 0.23 | | |
| 1 | 1.76 ± 0.31 | 1.58 ± 0.35 | 10.23 | |
| 3 | 4.37 ± 0.70 | 3.33 ± 0.99 | 36.16 | 1.38 |
| 6 | 8.56 ± 0.94 | 2.81 ± 1.54 | 62.97 | 8.28 |
| 12 | 16.24 ± 1.24 | 2.61 ± 1.34 | 83.90 | 9.01 |
| 24 | 32.68 ± 1.22 | 2.55 ± 1.41 | 92.20 | 10.20 |

For samples containing fibers with 500 and 1000 μm diameter, the measured $k_2$ values were close to those observed in fiber-free ones (Fig. 14). Conversely, by decreasing D, a significant improvement in the number of paths was noticed, indicating that a threshold diameter should be selected to induce the generation of well-distributed permeable paths in the resulting microstructure. Moreover, one should keep in mind that the design of the most suitable fiber diameter, length and content may also consider the rheological drawbacks that might arise from their use [15,77].

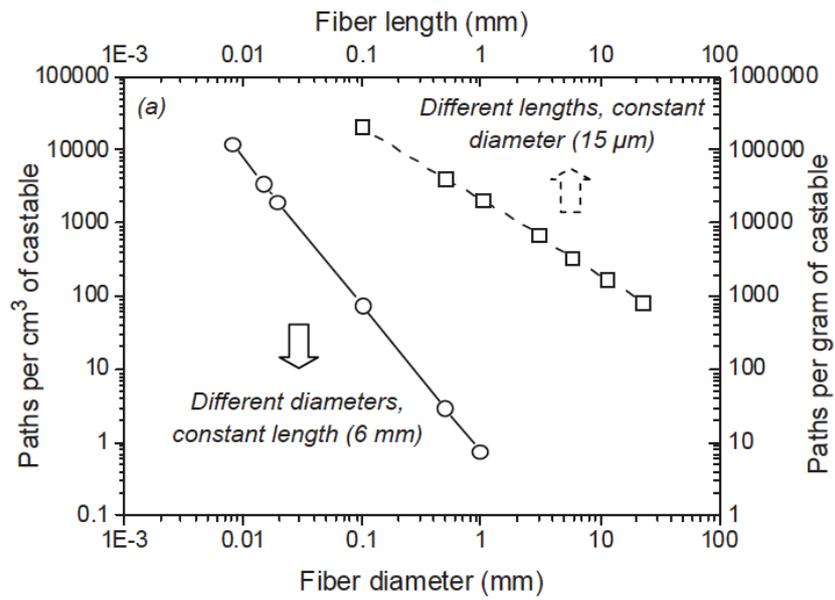

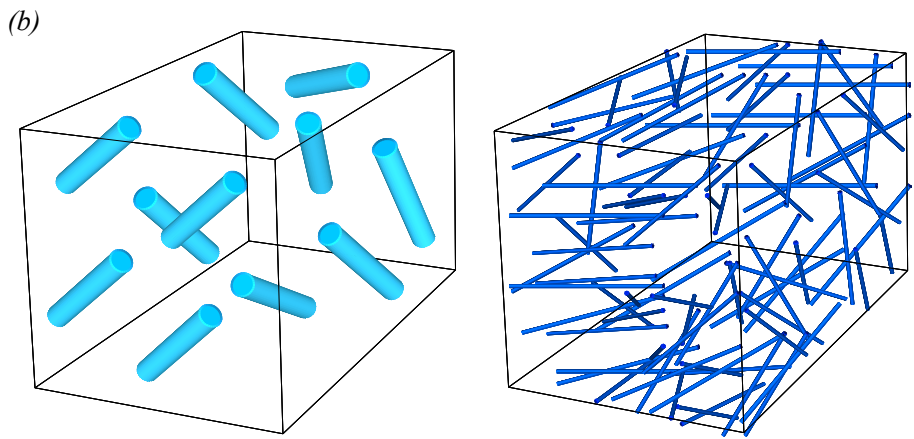

Figure 13: (a) Number of paths formed in the castables' structure after fibers' burnout as a function of the diameter (fixed length = 6 mm) and length (constant diameter = 15 μm) of these additives, considering the same added amount (0.36 vol%). (b) Sketch of the thick and thin fiber distribution in the castables' structure (for the same length and volumetric content) [75].

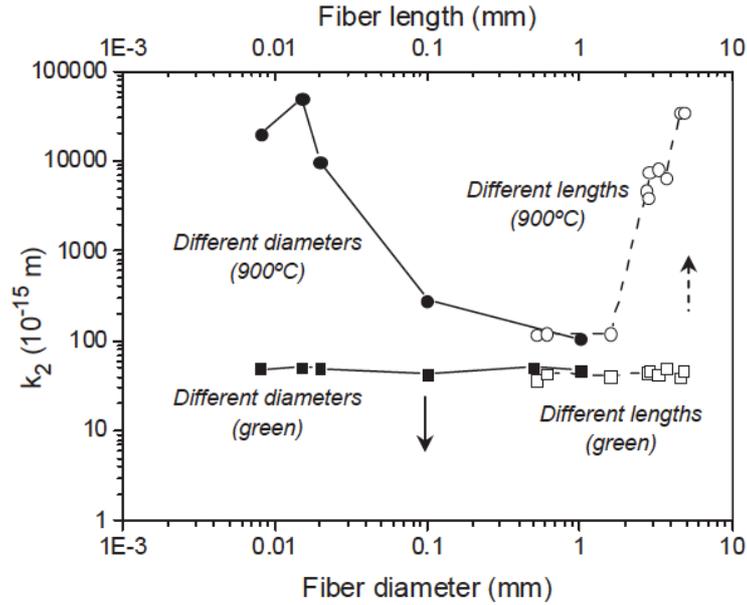

Figure 14: Permeability values ($k_2$) for green and fired (900°C) castable samples containing fibers with different diameters and lengths, for a constant added amount (0.36 vol%) [75].

*2.2. Metallic additives applied for improving the gas permeability of refractories*

It is common practice to add metallic additives (Al, Mg, Si and alloys) to refractory formulations as such materials may mainly prevent carbon oxidation, improve their spalling resistance and mechanical strength. They are incorporated into the compositions as powders and fibers and tend to react with water, generating a variety of compounds (metal-based oxides / hydrates) and hydrogen ($H_2$). The performance of such additives is influenced by various parameters (i.e., amount added, particle size, other raw materials contained in the mixture, curing temperature, etc.) always being aware of the dangerous shortcomings associated with the explosive nature of $H_2$ [76,78–80].

Aluminum powder is the most applied metallic additive to improve the permeability of refractory castables and inhibit their explosion during drying [78,79,81–85]. Its action is based on the reactions presented in Eq. 1-3 (which are mainly induced in high-alkaline conditions [86,87]) and the $H_2$ formation is able to generate permeable paths during the samples' curing step with the gas migration towards the refractory's surface. The explosion resistance of Al-containing castables depends on the elapsed time for gas generation and the mixtures' workability. For instance, if hydrogen-gas evolves when the castable is still fluid, no improvement in spalling resistance should take place as the final microstructure may not retain the memory of the permeable paths created during the migration of bubbles to the samples surface. On the other hand, if the hardening of the mixtures occurs prior to Al reaction, the evolved gas may be trapped temporarily and pressurized inside the rigid structure, which precludes helpful increase in permeability and causes occasional mechanical failures and/or swelling of the product [13]. The adjustment of this optimum time for the Al-water reaction may be achieved by controlling the castable hardening by choosing a suitable curing temperature or using accelerating or retarding additives [13, 79].

Another important aspect is that such transformations are exothermic, thus the heat release derived from Al hydration may also affect the binders' reactions during the castables' curing step.

$$2\ Al + 6\ H_2O \rightarrow 2\ Al(OH)_3 + 3\ H_2 \tag{1}$$

$$2\ Al + 4\ H_2O \rightarrow 2\ AlOOH + 3\ H_2 \tag{2}$$

$$2\ Al + 3\ H_2O \rightarrow Al_2O_{3(gel)} + 3\ H_2 \tag{3}$$

As illustrated in Fig. 15, the hydrogen generation in high-alumina calcium aluminate cement (CAC)-bonded castables can be divided into four well-defined stages. *Stage 1* depends on temperature and oxygen concentration, and it is related to the induction time. After that, a complex film with non-uniform morphology and composition begins to grow, forming an amorphous hydroxide layer (boehmite) on the aluminum surface (*Stage 2*). *Stage 3* corresponds to the greatest changes in the pressure profile and in the gelatinous boehmite film with the subsequent growth of bayerite or well-crystallized boehmite phases in this region. By increasing the Al content, the latter stage started slightly earlier leading to a remarkably higher reaction rate, H$_2$ generation and heat release (sample's temperature was 5°C higher for the 0.3 wt.% Al-containing composition when compared with the aluminum-free one, Fig. 15). However, the addition of a higher amount of Al powder did not result in a proportional increase in the hydrogen volume evolved from the evaluated castables. It was reported that such behavior might be related to: *(i)* the partial passivation of uncoated aluminum prior to its addition to the mixtures; *(ii)* the lower relative amount of water and oxygen available for the hydration reactions (Eq. 1-3) with the Al content increase; and *(iii)* the higher concentration of gelatinous products, which prevented the reactant diffusion to the aluminum surface, leading to an earlier suppression of the reaction in Stage 3 [81]. Besides that, using coated or uncoated Al powder will also influence the $H_2$ generation. In general, due to safety reasons, the selected metal is coated with a mineral oil, to prevent the formation of Al aerosols during conveying, storage and handling, reducing the combustion likelihood.

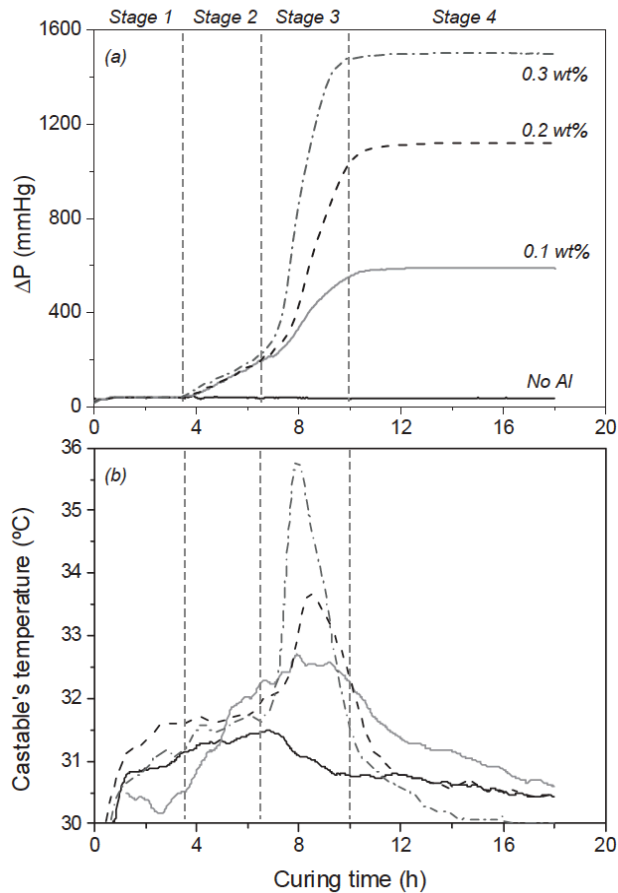

Figure 15: Stages of hydrogen generation as a function of the curing time and aluminum (uncoated metal powder) content incorporated into aluminous refractory compositions. (a) Pressure increase and (b) castable's temperature profiles. The curing process was carried out at 30°C [81].

When analyzing the permeability level and drying behavior of Al-containing CAC-bonded castables (Fig. 16a), it was observed that the generation of permeable paths was not necessarily enhanced by increasing the aluminum content. This performance can be explained by the fact that Al reaction with water (Eq. 1-3) not only resulted in hydrogen generation, but also hydroxides and alumina gel, which fills in the matrix interstices and reduces the overall $k_1$ and $k_2$ values [33].

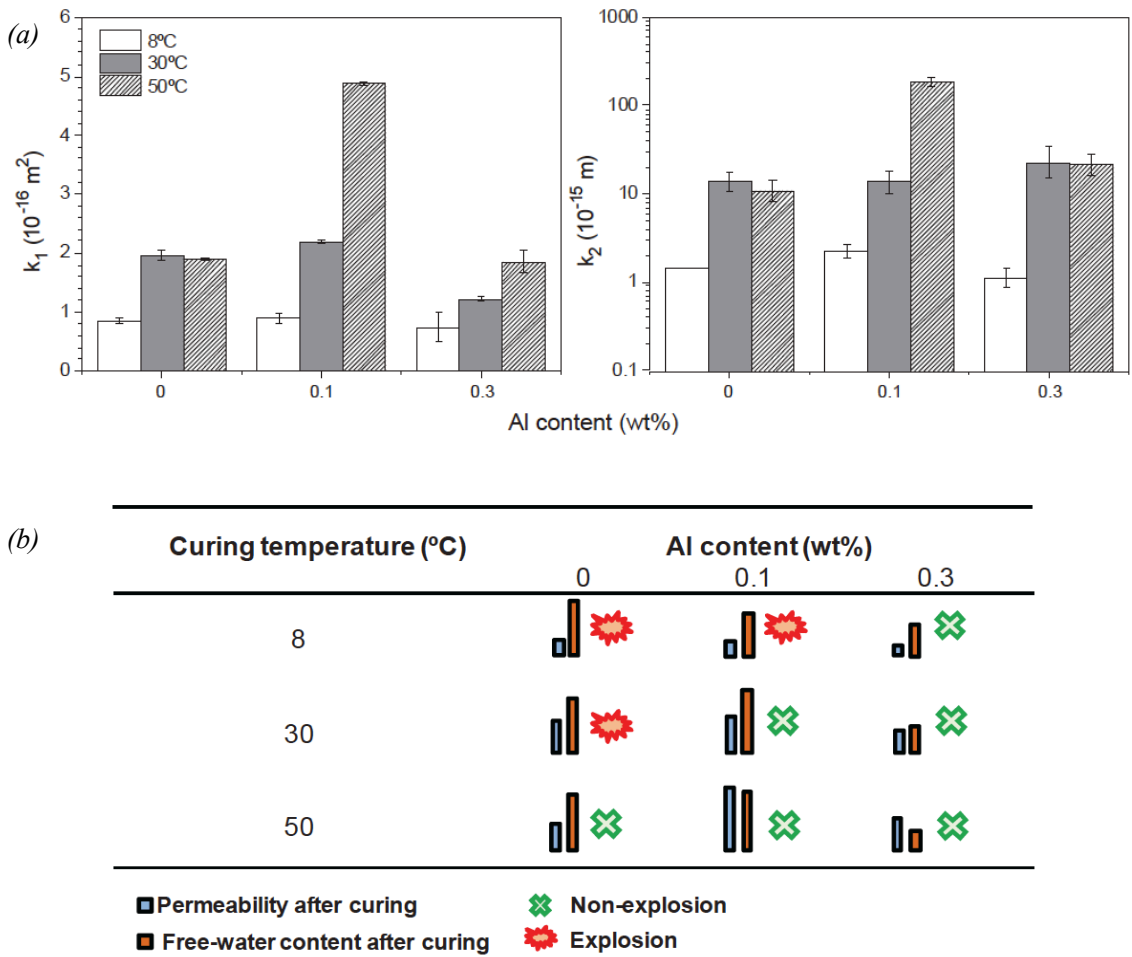

Figure 16: Influence of aluminum content on the (a) permeability and (b) explosive spalling likelihood of high-alumina castable bonded with 2 wt.% of CAC and containing distinct aluminum powder amounts (0.1 or 0.3 wt.%) and cured at different temperatures (8, 30 or 50°C) [33].

Furthermore, the additive's particle size and the selected curing temperature affect the reaction kinetics on the aluminum surface [33,84,85]. It is desired to have a balance between the gas evolution rate and the castable's hardening, as if gas is evolved during the mixture's mixing and placing, an excessive amount of water can be withdrawn prior to the suitable binder hydration and setting. On the other hand, if the hydrogen release is delayed (after hardening), the formation of permeable paths is spoiled and the system becomes pressurized, leading to the samples' swelling. Fig. 16b points out that only the castable composition containing 0.3 wt.% of Al was able to withstand a heating rate of 18°C/min after a curing step at 8°C. This indicates that the lowest temperature led to a less permeable refractory. Additionally, the test also pointed out that by increasing the amount of Al, the content of free-water released during drying was lower. This is an advantage of this additive compared to organic fibers, as permeable paths are generated earlier for the former and do not require higher temperatures to develop this structure [33,45].

Although most of the works presented in the literature are focused on evaluating Al powder, the use of aluminum fibers has also been investigated. According to Li and colleagues [78], this additive effectively improved the $Al_2O_3$-SiC-C castables' explosion resistance because, after its reaction with water, various pores could be formed at the fiber-matrix interfaces and the water vapor and $H_2$ accumulated in these areas generated a pressure level that induced a further shrinkage of the fibers. Consequently, the permeability of the refractory structure was enhanced, and novel paths could be formed with the release of gases (Fig. 17a).

Besides that, unlike the powder, residual Al from the fiber could still be found in the microstructure after its reaction with water. This remaining metal was molten during heating and even aluminum vapor could be generated at higher temperatures. The deposition of volatilized Al on the pore walls and its subsequent oxidation formed a dense alumina layer in the regions close to the original fibers. Consequently, some pores left by this metallic additive could still be identified in the refractory microstructure after firing at 1500°C (Fig. 17b). Considering these aspects, the authors reported that when the aluminum fiber exceeded the amount of 0.2 wt.%, the cold mechanical strength of the castables decreased due to the formation of a high volume of interconnected pores. Therefore, the best performance (explosion resistance, volume stability and mechanical strength) was observed for the compositions containing 0.10-0.20 wt.% of aluminum fibers [78].

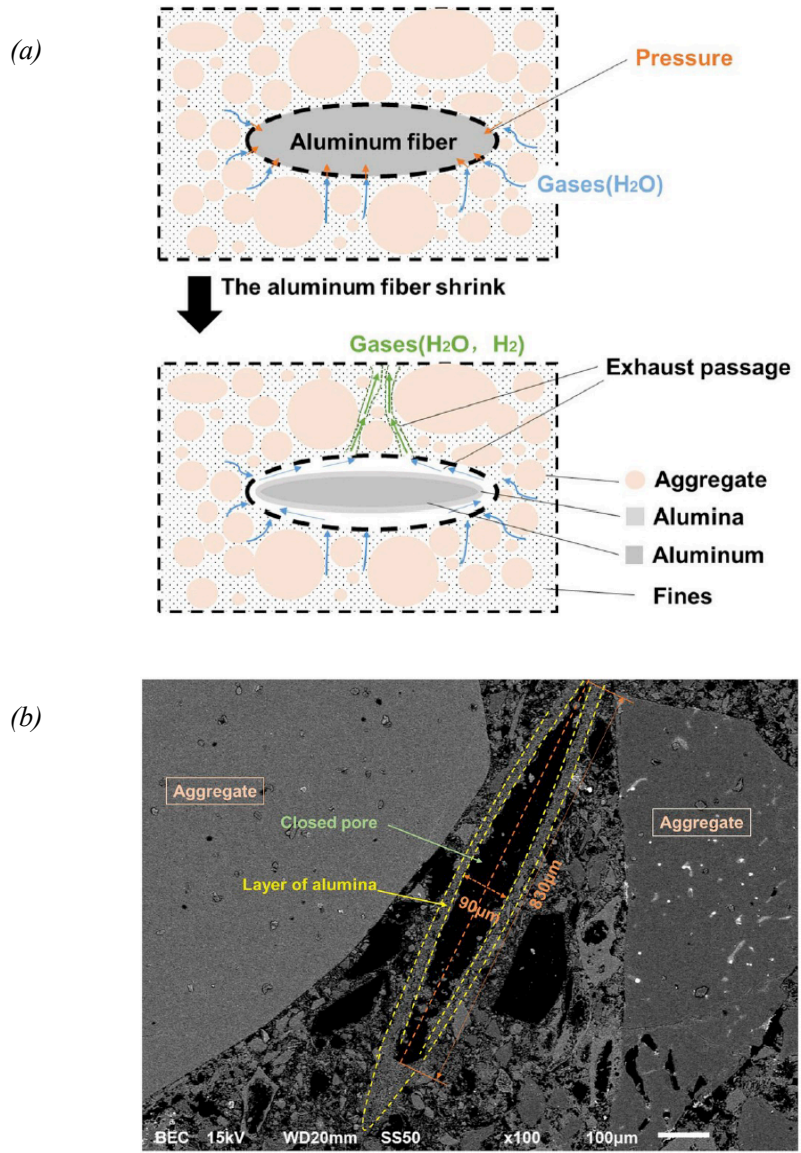

Figure 17: (a) Sketch of the aluminum fiber shrinkage and the formation of permeable paths when the generated gas escape from the resulting structure. (b) SEM image of the closed pore left by aluminum fiber in the $Al_2O_3$-SiC-C castable microstructure after firing at 1500°C. Amount of added Al fiber = 0.20 wt.% [78].

*2.3. Permeability enhancing active compound*

There has been increasing interest in different technologies to adjust and obtain refractories with sufficient permeable microstructure, aiming to release water at a drying temperature as low as possible. Thus, although the incorporation of organic fibers and aluminum powder is largely applied, some studies have pointed out a novel concept to increase the gas permeability for cement bonded castables based on the development of an engineered product (comprised by a mixture of organic and inorganic materials, such as $Al_2O_3$ = 39-43 wt.%, CaO = 12-15 wt.%, MgO 16-20 wt.% and others [34]), which induce the generation of connected pores at the beginning of the heating stage [16,20,35,58,88].

This additive (also called PEAC = permeability enhancing active compound) can be easily dosed in the castable dry-mix and it has the ability to modify the usual CAC hydration process, leading to the formation of gelatinous hydrated phases (Fig. 18) that show reduced thermal stability and, consequently, decompose around 80-120°C [16,88]. Thus, the shrinkage, cracking and degradation of the amorphous hydrates during heating are responsible for increasing the castables' permeability.

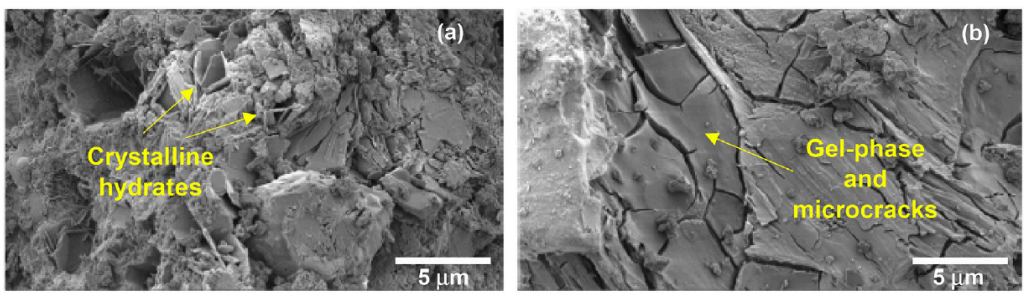

Figure 18: SEM images of high-alumina castable's matrix containing (a) 15 wt.% of calcium aluminate cement or (b) 12.5 wt.% of cement and 2.5 wt.% of PEAC. The presence of crystalline hydrates is pointed out in the additive-free composition, whereas gel-phase and microcracks could be observed when the additive was incorporated into the mixture [16]

As an example, Fig. 19 shows a comparison of the permeability level obtained for different alumina-based castable systems containing CAC or colloidal silica as binders. The $k_1$ results, measured after drying the samples at 110°C, for the standard cement (additive-free) formulation are very low due to the precipitation of the hydrates during curing and the resultant well-packed structure. A similar behavior was also observed for the composition containing 0.1 wt.% of PP fibers, as at this condition this additive has not yet melted. Conversely, the efficiency of the PEAC was already identified at 110°C, which contributed to a major improvement of the samples' permeability with values even higher than the ones collected for a colloidal silica-bonded material [20].

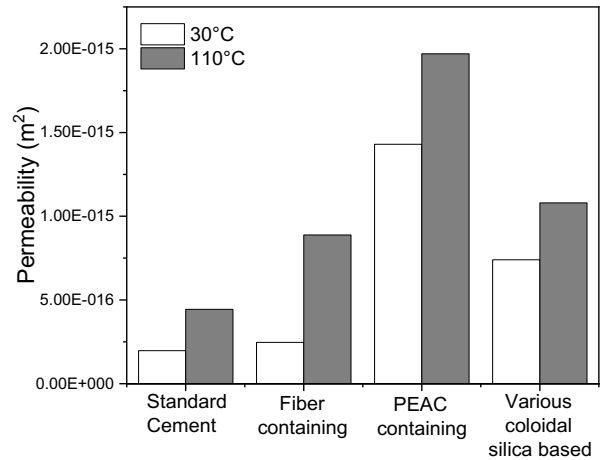

Figure 19: Permeability values ($k_1$) obtained for alumina-based CAC-bonded castables with no drying additive (standard cement), 0.1 wt.% of polypropylene fibers (fiber containing), 2.5 wt.% of a permeability enhancing active compound (PEAC containing). A colloidal silica bonded composition containing 2.5 wt.% of silica fume (various colloidal silica based) was also analyzed [20].

When analyzing pre-fired samples (300°C, Fig. 19), all evaluated castables had an improvement in the $k_1$ results and the most significant raise of the permeability was detected for the fiber-containing composition. Despite this performance, one must keep in mind that explosive spalling can take place already at 150°C, in which case a high permeability only at 300°C may be too late. Therefore, an important aspect when drying out a refractory castable is the temperature at which the water will evaporate and withdraw from the material, as this has an impact on both the dry-out kinetics, as well as the safety features during heating.

Although the PEAC has been initially designed for CAC-containing systems, Fig. 20 points out its positive influence in anticipating the water withdrawn and increasing the drying rate of high-alumina refractories bonded with hydratable alumina (AB) or two different cements (Secar 71 or Secar 51). The 6AB samples (additive-free) exploded during the thermogravimetric tests carried out with 20°C/min due to their reduced permeability (Fig. 20a), whereas the ones containing this active compound (3.5AB-AC) were able to withstand the pressure build-up and release a high amount of water at temperatures around 224°C (Fig. 20b). Major changes in the mass loss rate could also be detected for the cement bonded castables and the steam release took place at temperatures as low as 128°C [34].

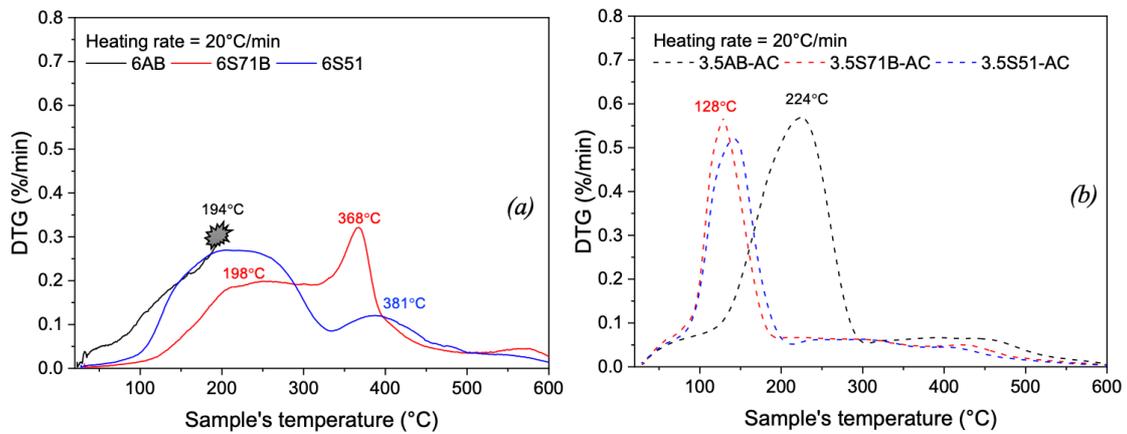

Figure 20: First derivative of the mass loss curve for the high-alumina castables bonded with hydratable alumina (AB) or calcium aluminate cements (Secar 71 or Secar 51) under a heating rate of 20 °C/min. The additive-free samples (a) and the ones containing 2.5 wt.% of the permeability enhancing active compound (b) were cured at 30°C for 1 day in a humid environment before the TG measurements [34].

This behavior is explained by the changes of the hydration reactions and the formation of gel-like hydrates in the designed castables due to the action of this PEAC. Fig. 21 shows the flow rate measured at room temperature for the additive-free and active compound (AC)-containing compositions, highlighting the benefits of this additive in enhancing the samples' permeability. As observed, the refractory bonded with hydratable alumina (3.5AB-AC) was the one that presented the greatest improvement of its flow rate (Fig. 21b), indicating that this compound can be used in different formulations and still result in safer drying conditions.

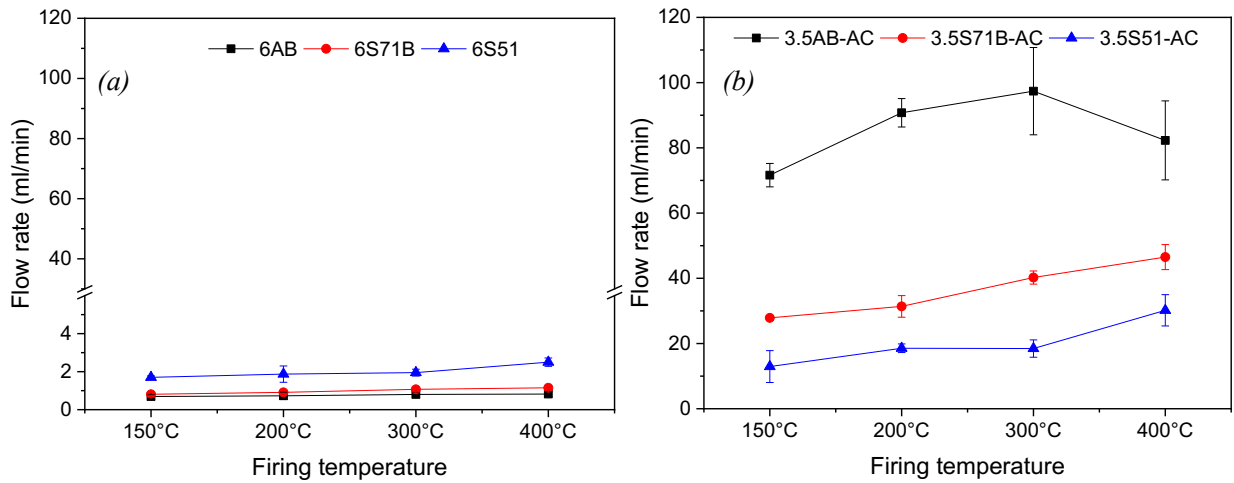

Figure 21: Room temperature air flow rate (ml/min) measured during the permeability tests of the castable samples containing (a) 0 wt.% or (b) 2.5 wt.% of the permeability enhancing active compound (AC) for a fixed inlet air pressure of 0.11 MPa. The specimens were cured at 30°C/24h, dried at 110°C/24h and fired between 150-400°C/5h [34].

Nevertheless, there are also some drawbacks associated with using this permeability enhancing active compound. Lower flowability, faster setting and reduced green mechanical strength are some of the main identified side effects for castable compositions [20,34,89]. Fig. 22 highlights the drop in the cold flexural strength obtained for alumina-based castables bonded with CAC or hydratable alumina (HA) and containing this active compound (named here as MP). According to these results, the additive-free and polymeric fiber (ED)-containing refractories presented higher mechanical strength in all tested conditions (30-400°C), whereas the precipitation of gel-like phases and the increase in the number of permeable paths in the samples containing PEAC led to a drop in the values of this property.

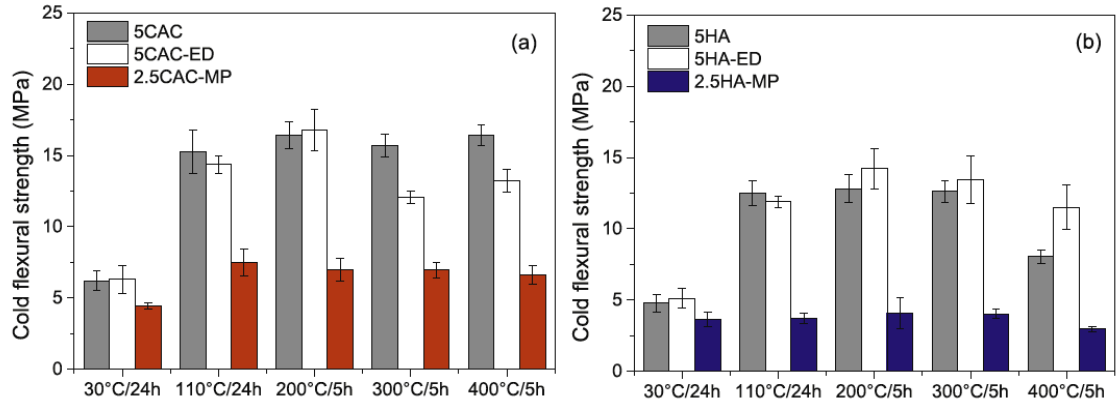

Figure 22: Cold flexural strength evolution of (a) calcium aluminate cement (CAC) or (b) hydratable alumina (HA)-bonded castables obtained after curing (30°C/24h), drying (110°C/24h) and firing (200-400°C/5h). ED = polymeric fibers and MP = permeability enhancing active compound [89].

*2.4. Silica-based additives*

Novel additives have been developed to enhance the permeability of castables, without affecting, to a greater extent, their green mechanical strength [34,90–92]. This is the case of a silica-based drying agent (SioxX-Mag), which was designed for magnesia monolithics presenting microsilica in their matrix fraction. According to some studies [38,93,94], adding 2 wt.% of this product to MgO-based castables containing 6 wt.% of $SiO_2$ resulted in compositions with improved flowability and mechanical strength. Besides that, crack-free large pieces could be produced, presenting higher explosion resistance.

Silica-based compounds are also efficient additives to control MgO hydration in refractory systems [95–98], because their dissolution in basic pH aqueous medium, gives rise to silicilic acid ($HSiO_3^-$), which favors the reaction of the latter with $Mg(OH)_2$ initially formed on the magnesia surface. Consequently, amorphous hydrated magnesium silicates (M-S-H gels) of very low solubility in water and with distinct stoichiometry, depending on the MgO and $SiO_2$ molar ratio (M/S), are formed on the magnesia particles' surface, inhibiting its further interaction with water [96,98,99]. Therefore, crack formation in the consolidated castables can be minimized with the reduction of brucite content generated in the matrix fraction of the microstructure.

Recently, Santos Jr. et al. [92] evaluated the incorporation of SioxX-Mag into $Al_2O_3$-MgO castables and confirmed the enhanced drying behavior of a composition containing 1 wt.% of this additive when compared to a refractory without silica. Fig. 23 shows the mass loss and the drying rate of such castables up to 600°C. The anti-hydrating effect derived from the silica-based additive was pointed out as the main aspect responsible for the higher explosion resistance of composition D. On the other hand, castable B (reference) exploded approximately at 450°C due to the pressurization of the structure associated with the decomposition of magnesium hydroxide.

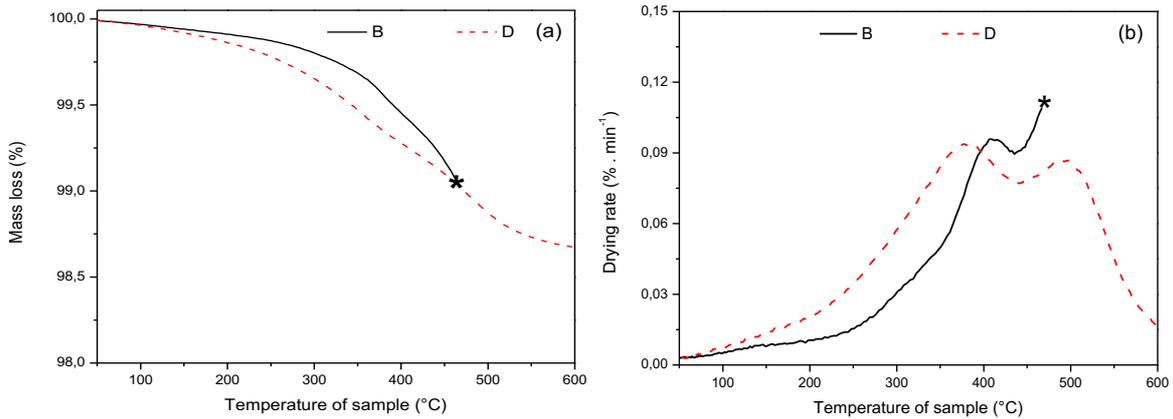

Fig. 23: Mass loss (a) and drying rate (b) of $Al_2O_3$-MgO compositions B (without silica) and D (with 1.0 wt.% of SioxX-Mag) obtained for thermal treatments carried out with heating rate equal to 20°C/min. The analysed samples were previously cured at 50°C/24h and dried at 110°C/24h. The asterisk (*) indicates the explosion of sample B [92].

Despite this positive effect confirmed by thermogravimetric tests, no permeability measurements have been reported in the literature for microsilica-gel bonded refractories [40,92–94]. Thus, additional investigations are still required to better understand how the silica addition might influence the formation of permeable paths in the resulting microstructure of some refractory systems.

*2.5. Chelating agent*

Among various organic additives suggested to improve the properties of cured magnesia containing compositions, aluminum lactate (aluminum salt of 2-hydroxypropanoic acid) is an interesting option as it can act as a chelating agent of $Mg^{2+}$ ions and induce the generation of hydrotalcite-like compounds [$Mg_xAl_y(OH)_{2x+2y}$]($CO_3$)$_{y/2}$.$nH_2O$] in such systems, which might lead to crack-free refractory pieces [37,100–104]. Previous works [105,106] also identified an additional benefit by incorporating this organic compound into magnesia-containing monolithic products. According to these studies, using a small amount of aluminum lactate in the designed compositions could favor the development of refractories with enhanced green mechanical strength, higher permeability and greater explosion resistance during their first heating treatment [105,106].

It is accepted that hydrotalcite-like phases might act preventing the castables' explosion because these compounds present crystalline lamellas (comprised by $Mg(OH)_2$ and $Al(OH)_3$ layers) separated by water molecules. Furthermore, some anions ($OH^-$ and $CO_3^{-2}$) can also be accommodated between these hydroxide lamellas to neutralize the electric charges. When the castables containing these phases are heated, their thermal decomposition will take place in various steps [100,107,108]: *(i)* removal of free or adsorbed water (< 100°C), *(ii)* elimination of the interlamellar structural water (100-200°C), and *(iii)* the simultaneous dehydroxylation and decarbonation of the layered double hydroxide framework (300-400°C). As a result, a more porous microstructure should be formed at lower temperatures (< 400°C) when compared to MgO-bonded compositions (which present only $Mg(OH)_2$ as the hydrated phase).

To illustrate this, Fig. 24 shows a comparison of the mass loss and DTG profiles obtained for MgO-bonded castables containing no additives (reference), polymeric fibers (PF), an organic aluminum salt (OAS = aluminum lactate), silica-based additive (SM) and the PEAC (MP) mentioned in Section 2.3. The evaluated samples were only cured for 24 h before testing, and therefore a large amount of water was still present in the microstructure during their first heating step. This aspect combined with the fast-heating rate (20°C/min) and most likely low permeability of the $Al_2O_3$-MgO castable shifted the water evaporation and ebullition to higher temperatures, resulting in higher steam pore pressure and greater spalling risk. The polymeric fibers did not prevent RefM-PF explosion during fast heating (20°C/min) because, despite their low melting point (~104°C), this material acted enhancing the refractory permeability mainly after its full decomposition (> 300°C) [34]. The incorporation of only 1 wt.% of the silica-based additive (SM) to the tested composition was not enough to inhibit the samples explosion in the evaluated condition. This behavior was related to the fact that the amorphous hydrated compounds derived from SM and $Mg(OH)_2$ reaction could also fill in the pores of the castable's structure, which significantly reduced its permeability and, consequently, the steam withdrawal during heating.

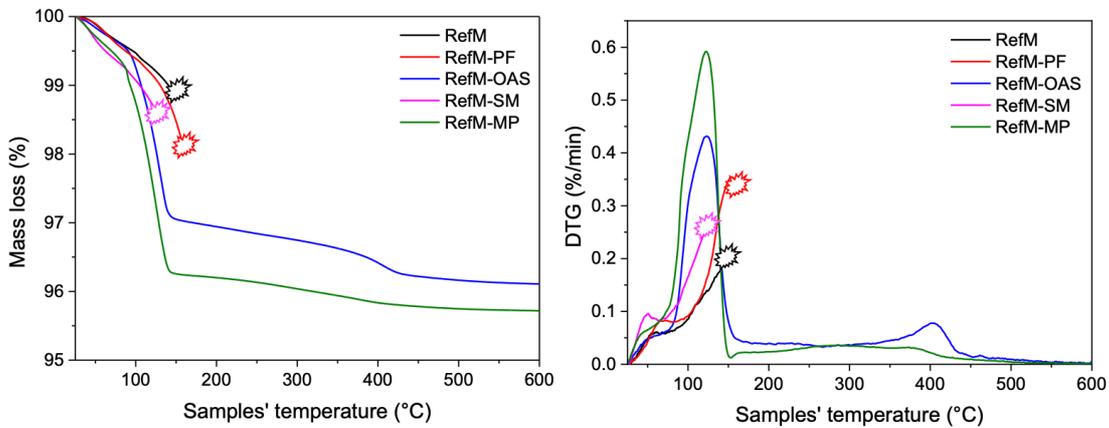

Figure 24: (a) Mass loss (%) and (b) first derivative of the mass loss curve for the thermal treatments carried out with a heating rate of 20°C/min for the $Al_2O_3$-MgO castables. All samples were cured at 30°C for 1 day before the measurements. PF = polymeric fibers, OAS = aluminum lactate, SM = silica-based additive and MP = Refpac Mipore 20 (PEAC) [37].

Regarding the castables containing aluminum lactate (RefM-OAS) or PEAC (RefM-MP), both were able to expel a higher amount of water in the 80-123°C range (more intense peak pointed out in Fig. 24b), indicating that most likely the hydrotalcite-like phases derived from the action of these two additives allowed interlamellar water withdrawal at low temperatures or the gel-like phases to undergo decomposition at such conditions, giving rise to a great number of permeable paths in the resulting microstructure [34,109]. Additionally, another interesting aspect is that the RefM-OAS sample also presented a further broad decomposition range (peak around 402°C), which can be related to $Mg(OH)_2$, hydrotalcite-like phases and aluminum lactate decomposition [37].

Nevertheless, the amount of this organic salt should be adjusted when added to refractories containing different MgO sources. Depending on the magnesia content and reactivity, distinct quantities of aluminum lactate will be required to effectively limit $Mg(OH)_2$ generation and induce the formation of the hydrotalcite-like phases [$Mg_6Al_2(OH)_{16}(OH)_2.4.5H_2O$ / $Mg_6Al_2(OH)_{16}(CO_3).4H_2O$]. A recent investigation highlighted that different permeability levels and drying behavior could be observed for castables containing dead burnt (DBM), caustic (CM) or magnesia fumes (MF) and 0.5 or 1.0 wt.% of aluminum lactate (AL) [110]. Fig. 25 indicates that CM-1AL castables presented the lowest $k_1$ and $k_2$ values after drying or firing the samples in the 110-450°C temperature range when compared to the other compositions containing DBM or MF. However, the obtained permeability results were higher than the ones commonly observed for calcium aluminate cement (CAC) or hydratable alumina (HA)-bonded refractory systems [34,89]. Hence, AL addition to the designed compositions induced the generation of a greater number of permeable paths, which is a key aspect for safe and fast drying, as the castable's mechanical strength itself may not be enough to withstand the high thermomechanical stress.

Despite the good performance in MgO-bonded refractories, the incorporation of AL in cement or hydratable alumina-containing formulations is not a straightforward task, as the same drawbacks pointed out for PEAC (decrease in the flowability, setting time and green mechanical strength) could also be detected in this case. Thus, based on the studies available in the current literature, this organic salt is a good alternative to improve the drying performance of MgO-containing monolithics [37,105,106,110].

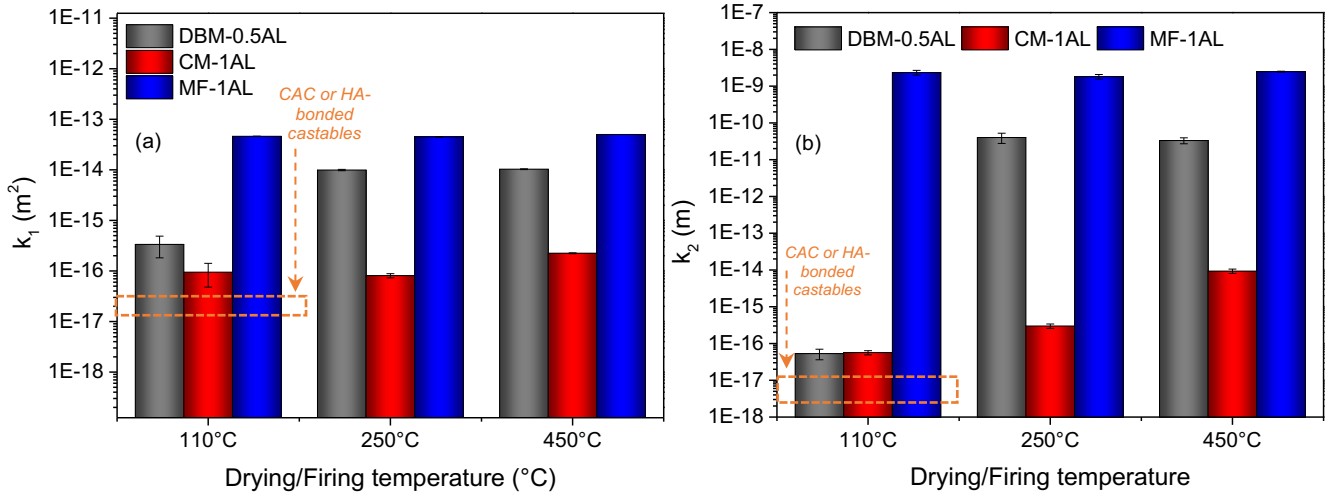

Figure 25: Permeability constants (a) $k_1$ and (b) $k_2$ measured for the dried and fired samples of the aluminum lactate-containing castables. DBM = dead burnt magnesia, CM = caustic magnesia and MF = magnesia fumes [110].

## 3. Design of drying schedules for different refractory compositions

As described in the first part of this review series [2], the dry-out procedure of refractory products can be divided into three main steps: evaporation, ebullition and decomposition of the hydrated phases. The schedules should be developed for the specific lining/unit to be dried out and various other aspects (binder type, amount of water to be released, permeability level, etc.) of the refractory material must also be considered. For castable linings, historically the procedures proposed are based on ramp (heating rates usually do not exceed 25-75°C/h) and dwell time (held at some selected temperatures for specific periods of time). Table 6 highlights some examples of heating-up schedules proposed by a refractory manufacturer [111].

As pointed out in many studies focused on the drying behavior of castables [26,32,50,58,112], the ebullition stage (110-300°C) is the most critical one and, aiming to minimize the steam entrapment within the ceramic lining structure and favor water withdrawal at the beginning of the heating process, it is of utmost importance to properly design and select the drying schedules to be used.

Thus, many efforts have been carried out to investigate the influence of distinct heating rates, the use of continuous heating and temperature plateaus as well as the effect of the castables' dimension on the early steam release from dense castables during their first thermal treatment. The following sections will present some examples and highlight important aspect required for designing safer and optimized drying schedules.

Table 6: Recommended heating up schedules of refractory castables as proposed by AGC Ceramics [111].

| **Heating-up schedules** | **Product category** | | |
| --- | --- | --- | --- |
| | **Low Cement Castables for Aluminum industries** | **Explosion-resistant Low Cement Castables** | **Conventional Castables** |
| From room temperature to 200°C | 50°C/h | 50°C/h | 50°C/h |
| Maintain a temperature of 200°C for certain hours calculated as per the instructions provided. | 1h for each 10 mm lining thickness. Example: for a lining with 200 mm thickness, the holding time at 200°C should be 20h. | 1h for each 20 mm lining thickness. | 1h for each 20 mm lining thickness. |
| From 200°C to 350°C | 25°C/h | 25°C/h | 50°C/h |
| Maintain a temperature of 350°C for certain hours calculated as per the instructions provided. | 1 h for each 10 mm lining thickness. | 1h for each 20 mm lining thickness. | 1h for each 20 mm lining thickness. |
| From 350°C to the operation temperature | 25°C/h | 50°C/h | 50°C/h |

*3.1. Heating rates' influence on the drying behavior*

Among the parameters to be chosen, the heating rate plays a crucial role as, although fast heating is desired to reduce the overall time required for the drying process, it might also induce an uncontrolled pressure build up in the refractories and favor the development of cracks or even the disintegration of the lining.

To point out the impact of this variable in the water release behavior of cured samples, Figure 26 shows the mass loss and DTG profiles of alumina-based CAC-bonded castables subjected to four different heating rates (0.9, 3.7, 6.4 and 10.4°C/min) during thermogravimetric tests (cylindrical samples with d = 40 mm and h = 40 mm) carried out up to 300°C. The time required for complete drying (to achieve 100% of mass loss, Fig. 26a) decreased with the applied rate, but an inversely proportional increase in the final temperature ($T_f$) to reach this condition was also observed (Fig. 26b). Hence, although faster heating (changing from 0.9 to 10.4°C/min) could favor a reduction of 87.5% in the overall drying time of the samples, the identified drawback was the increase of roughly 45% in $T_f$ [30].

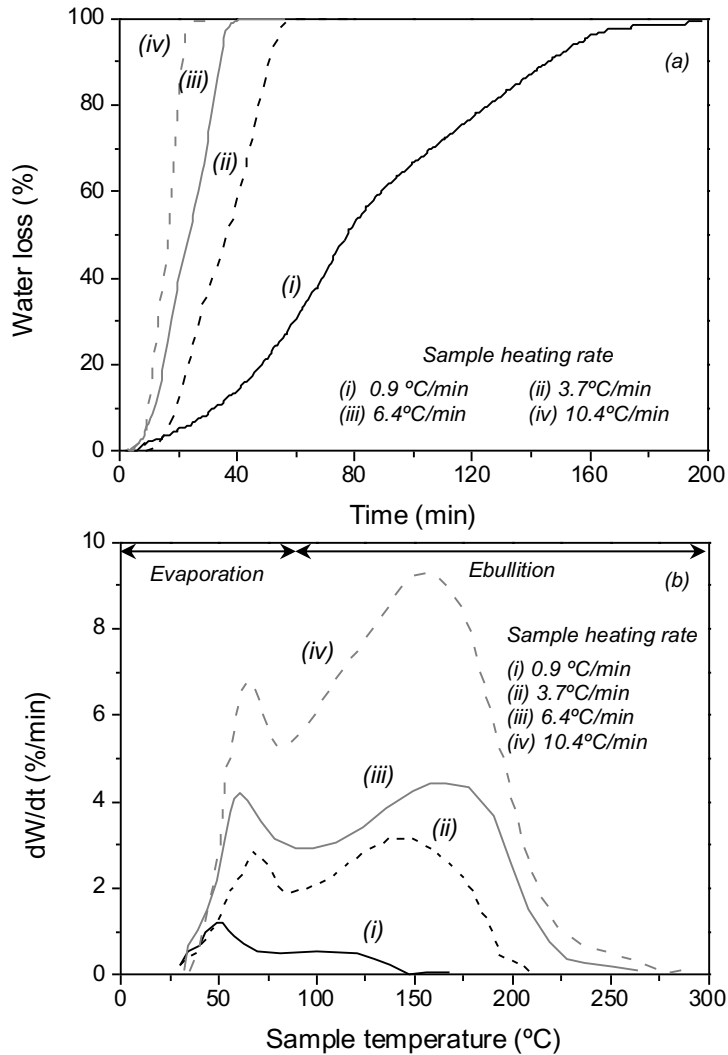

Figure 26: (a) Mass loss profiles as a function of time and (b) drying rate (DTG) of CAC-bonded alumina-based castables. The samples were subjected to distinct heating rates (0.9 to 10.4°C/min) [30].

Considering that the water vapor pressure scales exponentially with temperature as proposed by Antoine's equation (Eq. 4 and Fig. 27a, which is valid in the range of 0 to 374°C for liquid-vapor equilibrium condition [113]), higher heating rates tend to increase the castables' explosion spalling likelihood.

$$P_v = exp\left(A - \frac{B}{T+C}\right) \qquad (4)$$

where $P_v$ is given in Pascal, $T$ is the temperature (in Kelvin), and $A, B, C$ are empirical dimensionless constants (for water, A = 23.33; B = 3841.22; C = -45.00).

Fig. 26b points out two main peaks associated with the water evaporation and ebullition stages for the analyzed samples. When changing the rate from 3.7 to 10.4 °C/min, a shorter time was required to reach the ebullition temperature and, consequently, instead of 30 min a total of ~ 11 min was already sufficient to reach 110°C. The DTG peaks related to the evaporation process showed their maximum value close to 50-65°C for all tested conditions. Moreover, a five-fold increase in their intensity from the lowest to the highest rate could also be observed. The latter feature is important, because the castable's behavior during the ebullition stage will depend on the amount of water remaining in the structure at the boiling onset. According to the TG results, the increase in the sample's heating rate reduced the amount of liquid released between 20-110°C from 73 to 36%, raising the content to be removed as vapor above 110°C [30]. These results indicate that the evaporation performance is more dependent on the time rather than the actual applied heating rate. For this reason, it is highly recommended the use of "two-step schedule" or to apply a holding time step close to the water boiling point to maximize the amount of water withdrawal from the refractories in this first stage.

Regarding the drying rate above 110°C (ebullition, Fig. 26b), the peaks' temperature at their maximum value was shifted from 120 to 160°C when changing the heating rate from 0.9 to 10.4°C/min. Therefore, faster rates tend to enhance the likelihood of vapor pressure build up and to induce the castables' explosion to take place [30].

Aiming to comparatively evaluate the spalling likelihood during free-water release, Innocentini et al. [30] suggested the calculation of a parameter described as the potential risk of pore pressurization ($P_{risk}$, Eq. 5 and 6).

$$W_d \, (\%) = 100 \, x \left( \frac{M_0 - M}{M_f} \right) \tag{5}$$

$$P_{risk} = (W_{df} - W_d) P_v \tag{6}$$

where, $W_d$ is the overall water loss (based on the sample's dry weight), $M_0$ the initial mass, $M$ the instantaneous mass recorded at a certain time during the heating stage, $M_f$ the final (dry) mass of the sample. The parameter $W_{df}$ is calculated by Eq. 5 when $M = M_f$ and $P_v$ is obtained by Eq. 4.

Fig. 27 indicates the water vapor pressure evolution with temperature (according to Eq. 4) and presents the $P_{risk}$ values estimated for the refractory samples subjected to distinct heating rates. Knowing that the pre-fired castable tensile stress levels were around 1.4 - 3.0 MPa, the critical temperature interval ($\Delta T_c$, above which the refractory would most likely not be able to withstand the pressurization) was identified in Fig. 27a. As observed, the risk of pressurization increased with the applied heating rates, but no explosion took place because the maximum risk temperatures ($T_{max}$, Fig. 27b) were below $\Delta T_c$.

The $P_{risk}$ profiles are low at the beginning and end of the drying process, either due to initial low temperatures or as a consequence of the withdrawal of most of the free-water, respectively. This parameter is higher in the 100 and 160°C range (Fig. 27b) because the pressure values can be equal or greater than the castable's tensile stress. Although $P_{risk}$ might not be appropriate for assessing the vapor pressure derived from all types of hydrate decomposition (as Eq. 4 is only valid up to 374°C), it can be useful to compare the safety of various drying schedules for a given refractory product or to provide fundamentals for the optimization of the castable's microstructure [30]. Thus, $T_{max}$ can be reduced (and, consequently, the explosion risk decreased) when optimizing the permeability, or changing the heating schedule or favoring faster water release during evaporation.

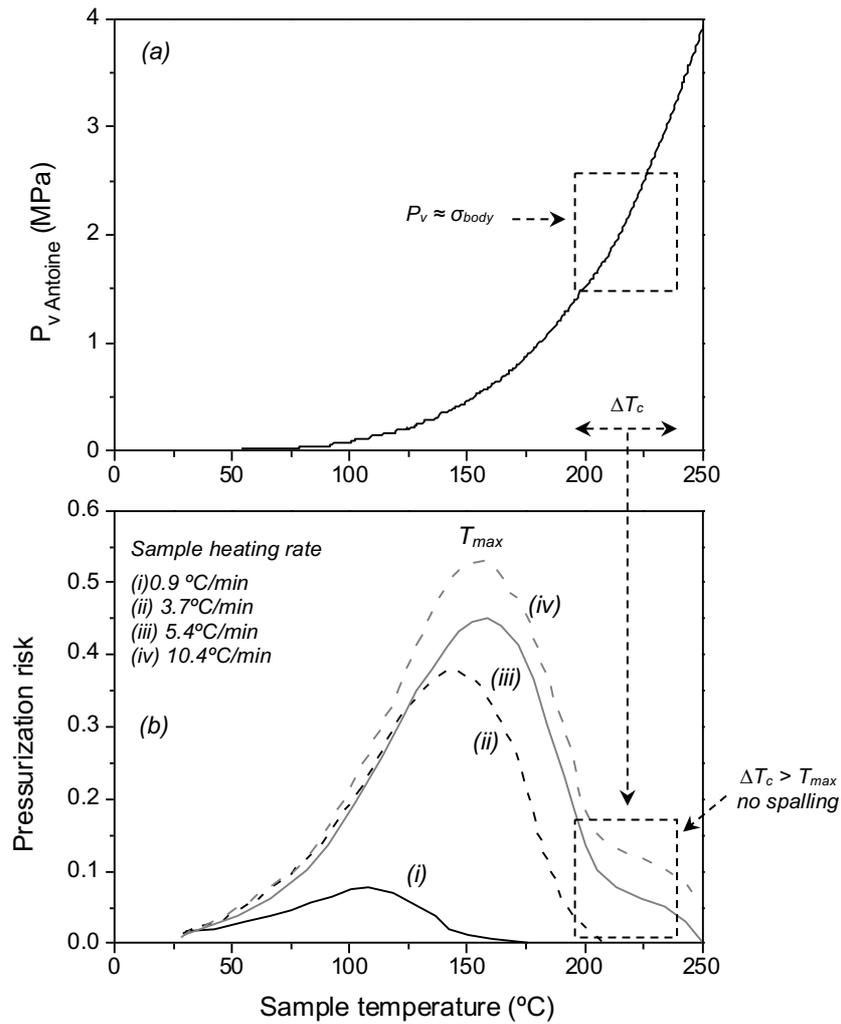

Figure 27: Maximum pressure build up inside the heated castables: (a) critical vapor pressure ($P_v$) calculated according to Antoine's equation and (b) pressurization risk ($P_{risk}$) estimated for various drying schedules (0.9 to 10.4°C/min) [30].

*3.2. Continuous heating or temperature plateau*

Distinct heat up procedures, based on ramp and/or holding steps, can be applied during the drying of refractories to maximize the $H_2O$ withdrawal at low temperatures (evaporation stage). Besides the water boiling point, one should also consider the type of hydrates and the refractories' microstructural features before defining at which temperature the holding steps could be applied.

To illustrate the effect of the drying schedule on the water loss of cured castable samples (cylinders with d = 40 mm and h = 40 mm), thermogravimetric analyses were carried out by some authors to evaluate the drying kinetics of high-alumina CAC-bonded compositions [31]. Two distinct thermal treatments were conducted: *(i)* furnace's temperature was increased from room temperature to 60°C in 20 min and the samples were maintained at this maximum temperature until 90% of the free-water content was removed (i.e., for more 340 min, which was named as *plateau* in Fig. 28); or *(ii)* furnace's temperature was slowly and gradually raised up to 60°C using the same overall time of 360 min considered for the first schedule, but without any holding time at 60°C (this procedure was called *ramp* in Fig. 28).

Distinct drying behavior could be observed for both analyzed conditions. Fig. 28a and 28b highlights that the faster heat up for the plateau procedure induced a more significant mass loss and a sharp *increasing rate period* (IRP) at the beginning of the drying process. This IRP condition took place in the castables due to the continuous heating procedure applied instead of the isothermal one. After reaching the peak at 50 min for the plateau treatment (Fig. 28b), the progressive retreat of the drying front into the structure pores led to the beginning of the *falling rate period* (FRP). On the other hand, a longer but less sharp IRP was verified for the ramp schedule, with a maximum peak detected around 170 min. Thus, if one desires to speed up water evaporation, the plateau method proved to be the most efficient procedure. Moreover, Fig. 28c shows that the end of the IRP in both schedules was followed by a transitory increase in the heating rate ($dT_s/dt$) measured at the sample's surface. This is an important parameter that can be used to control the dewatering profile of the ceramic lining during practical applications as the location of the drying front along the lining thickness can be monitored by these sharp increases in the temperature [31].

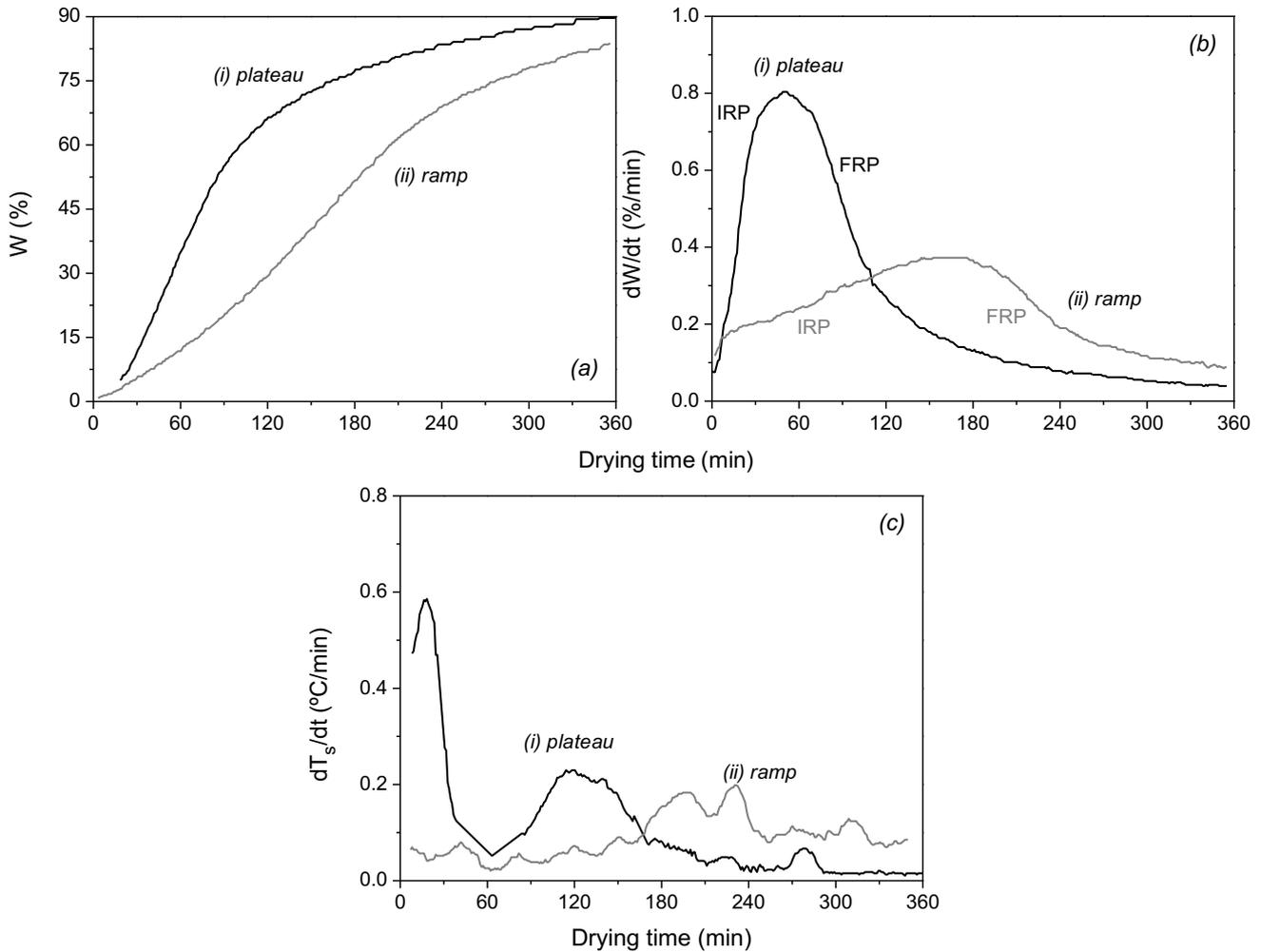

Figure 28: Thermogravimetric data of a drying process carried out at 60°C according to two different heating routines. (a) cumulative water loss (W) and (b) drying rate profiles of the evaluated samples. (c) Heating rate values at the sample's surface. All tests were conducted with a dry air stream inside the furnace. IRP = increasing rate period, FRP: falling rate period [31].

The evaporation rate can also be favored when increasing the maximum temperature of the environment, as it improves the driving force for the liquid capillary transport. For instance, Fig. 29 shows the same sort of experiments presented before (plateau versus ramp curves) but the selected temperature was 110°C instead of 60°C. In this case, the dry out was much faster and no signs of water ebullition were noticed, as the maximum samples' surface temperature for the plateau schedule (after 158 min) was 95°C. The main difference observed for the tests carried out at 110°C (Fig. 29b) was the change in the evaporation drying periods, with a more intense IRP drying rate and earlier transition to the FRP. The end of the IRP was detected by the higher thermal peaks indicated in Fig. 29c for both routines.

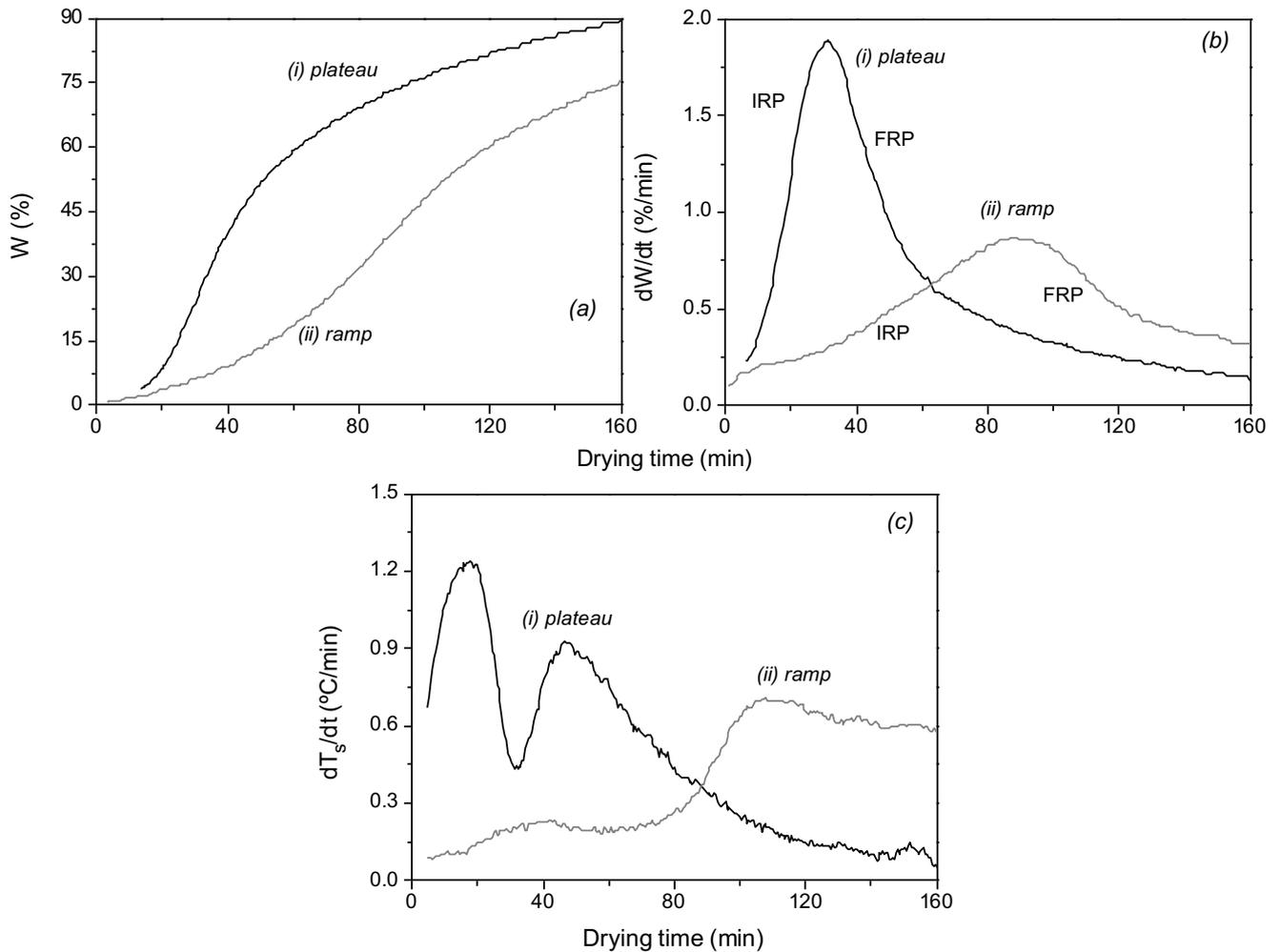

Figure 29: Thermogravimetric data of a drying process carried out at 110°C according to two different heating routines. (a) cumulative water loss (W) and (b) drying rate profiles of the evaluated samples. (c) Heating rate values at the sample's surface. All tests were conducted with a dry air stream inside the furnace. IRP = increasing rate period, FRP: falling rate period [31].

The following example highlights that the explosive spalling likelihood of cured refractories can be decreased by inducing a suitable removal of free-water in large castable samples. In such a case, 800 mm x 600 mm x 200 mm blocks (~300kg) of a microsilica-gel bonded refractory (without calcium aluminate cement or hydratable alumina) were prepared and dried at 160°C or 220°C before the drying tests. Two heat up schedules were selected, considering a holding time of 6 h or 10 h at 160°C (Schedule A) or 220°C (Schedule B), respectively (Fig. 30) [39].

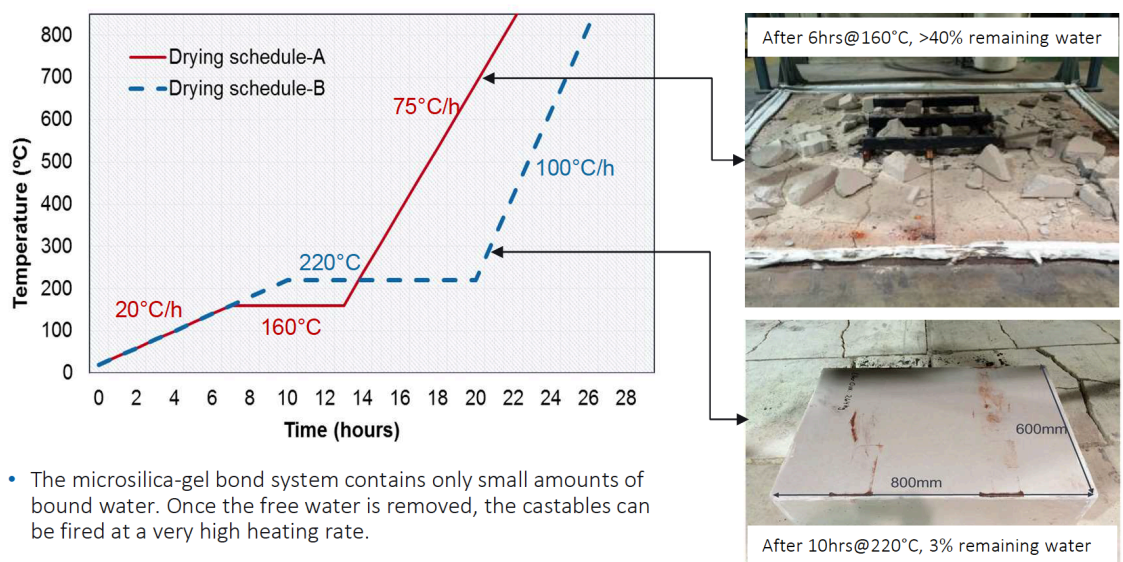

Figure 30: Heat up schedules applied for drying microsilica gel-bonded castables. Images of the 300 kg blocks obtained after the drying tests are also pointed out on the right side [39].

According to the obtained results, over 40% of water were still contained in the blocks kept at 160°C for 6 h, whereas the ones maintained at 220°C for 10 h presented a withdrawal of around 97% of water. Therefore, when the former specimens were heated from 160°C to 850°C at a heating rate of 75°C/h, the remaining water was responsible for increasing the vapor pressure in the core of the blocks, which eventually led to their explosion and disintegration (Fig. 30). On the other hand, no cracking or explosion was detected when following Schedule B due to the effective water removal during the holding time at 220°C.

After inducing the free-water release during the evaporation period, various ramp and plateau stages can be applied when designing a complete heat up schedule for the drying process of refractory linings. One should be aware that the selection of the heating rate should be adjusted to prevent the shift of the drying rate peaks to temperatures where the resultant water vapor pressure might exceed the castable's tensile strength (as discussed in Section 3.1). Furthermore, the selection of the temperature for the holding time should be based on the chemical composition of the refractories (i.e., considering the hydrates decomposition temperature).

*3.3. Impact of the castable's dimension on the spalling resistance*

In industrial applications, the dry out of refractory blocks with distinct dimensions takes place simultaneously resulting in a complex process, as each individual product behaves according to its particular features (i.e., permeability, thermal conductivity, amount of water, green mechanical strength, thickness, etc.). Consequently, the overall drying schedules should be designed considering the restrictions of the compositions with a higher spalling likelihood and complex geometry, even though other castables with higher permeability could withstand more aggressive dewatering conditions.

Besides that, the resistance to fluid flow in a well packed structure rises with the body's size because the vapor has a longer path to flow through the refractory before reaching the external surface. Hence, the castable's dimensions (thickness) must be accounted when planning the heat up step of ceramic linings (as pointed out in Table 6).

Laboratory tests were carried out by some authors [114] to illustrate the influence of the body's dimension on the dewatering process. Cylindrical samples (with diameter and height equal to 1.7, 2.1 or 5.2 cm) of alumina-based CAC-bonded castables were analyzed via thermogravimetric measurements up to 800°C. Fig. 31 shows that by increasing the samples' size, a decrease in their actual heating rate took place during a continuous heat up treatment carried out with 20°C/min (furnace's heating program). The higher displacement of the 5.2 cm sample profile confirmed its slower rate when compared with the ideal condition (y = x). Despite the lowest heating rate (which can be associated with the greater conductive resistance within the solid body and the absorption of part of the heat by the free-water), such larger pieces still presented explosive spalling.

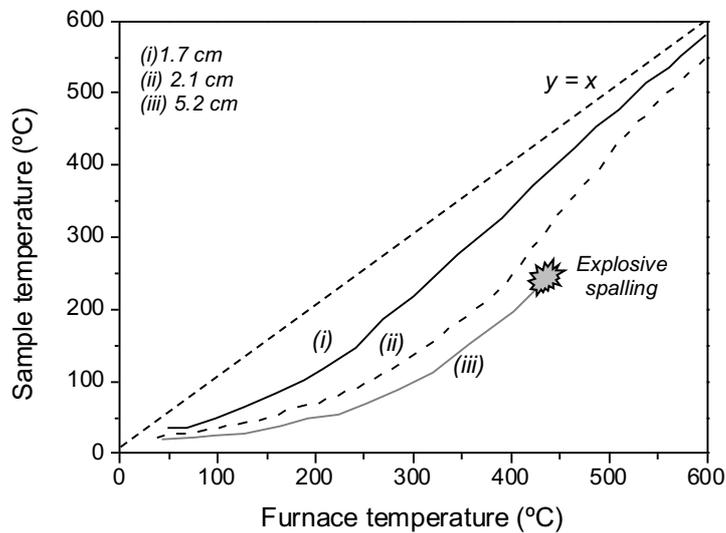

Fig. 31: Correlation between the sample's and the furnace's temperature for a high-alumina castable when testing distinct body sizes heated up to 800ºC under 20ºC/min [114].

According to the mass loss ($W$) profile and its first derivative ($dW/dt$) curve (Fig. 32a and 32b) as a function of time obtained for the tested castable, a shift of the water release was detected to longer times when increasing the sample's size. For example, 90% of water withdrawal was observed after only 15 min for the smallest sample (1.7 cm), whereas the 5.2 cm one showed less than 70% of water release after 25 min, when it exploded. When analyzing the drying rate with the samples' temperature (Fig. 32c), similar profiles (evaporation and ebullition peaks) were obtained regardless of the specimens' dimension, which indicates that these drying stages depend more strongly on the castable's temperature than on the environment one. Besides that, the intensities of both identified peaks decreased as the samples size scaled up, resulting in a broader ebullition peak temperature range. As a consequence, a fraction of the free-water was only released at higher temperatures, resulting in greater explosion risk due to steam pressurization in the castable's structure.

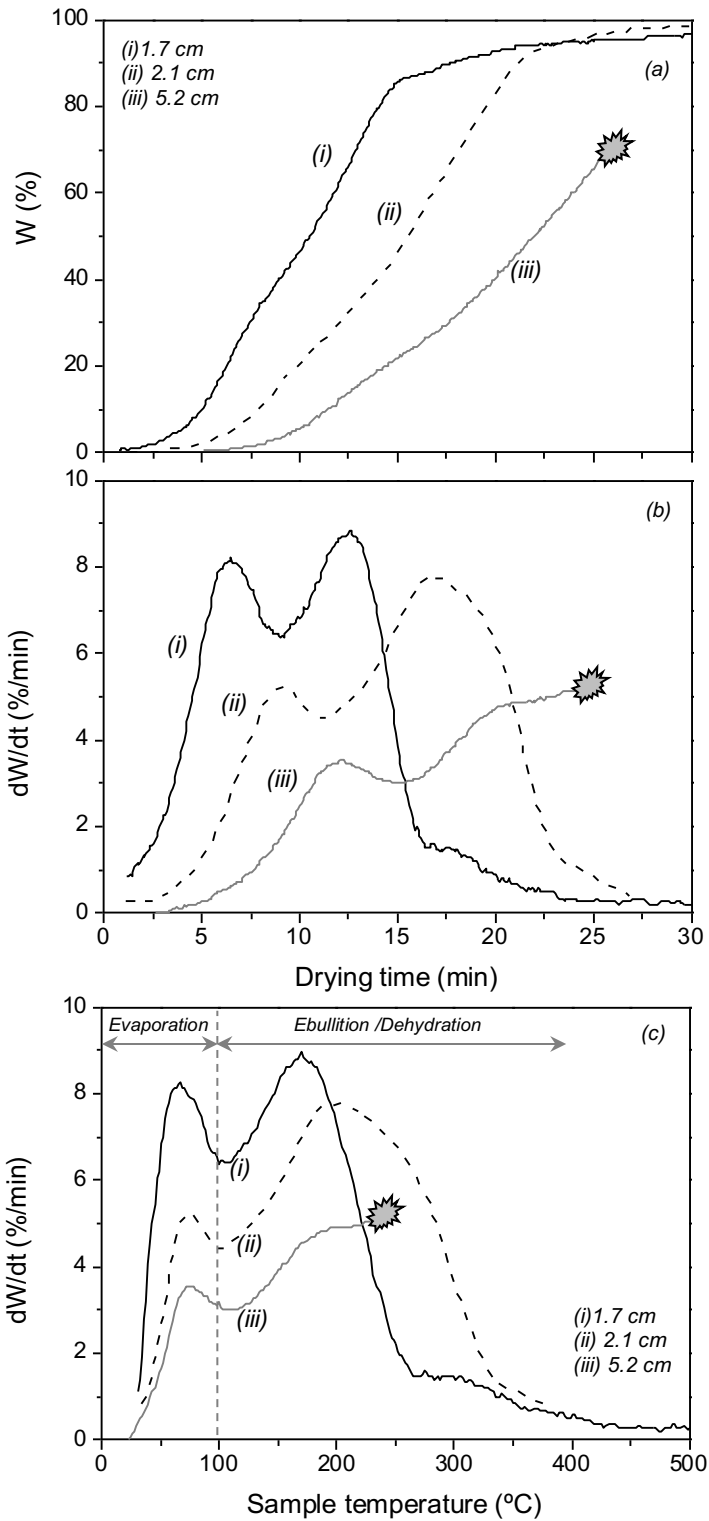

Fig. 32: Drying behavior of the high-alumina castables: (a) cumulative mass loss versus drying time, (b) drying rates for the distinct compositions as a function of time and (c) drying rates for the distinct compositions as a function of temperature [114].

Industrial-scale explosion resistance tests have also been carried out by some authors [39,40], using large blocks (as the ones shown in Fig. 33) of microsilica-gel bonded (non-cement, NCC) castable and low-cement containing compositions (LCC). The samples were cured at room temperature for 24 h, demolded and subjected to two heating schedules: NCC = 20 to 850°C at 50°C/h, and LCC = 20 to 850°C at 75°C/h. When investigating the spalling resistance of the NCC and LCC castables containing two different polymeric fibers, it was observed that the compositions containing 0.1 wt.% of a special fiber with lower melting point (100-120°C, named EMSIL-DRY) survived the industrial tests and no explosion took place. Additionally, a good correlation between previous lab-scale tests and the industrial ones could be obtained. Thus, Peng et al. [40] stated that lab-scale explosion measurements may still provide good guidance for industrial installation.

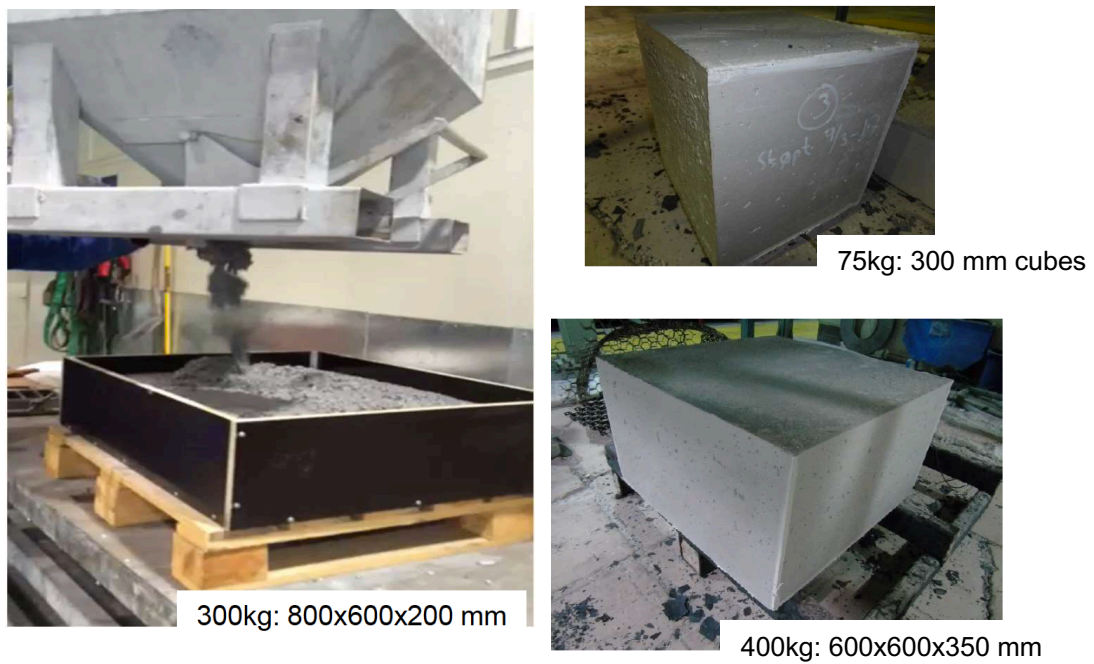

Figure 33: Large blocks of microsilica-gel-bonded castables prepared for the evaluation of their explosion resistance during heating up to 850°C with a rate of 50°C/h [39,40].

## 4. Final remarks

Although the evolution of refractory castables over recent decades has improved their performance, it has also posed some new challenges. One of them is the increasing complexity of the dry out step, due to the advances on the particle size distribution design and reduction in the lining's permeability. Thus, to ensure safe and fast heating schedules for refractory castables, some routes can be considered to improve the permeability of dense structures, such as the incorporation of additives (drying agents) into the refractory formulations to assist the generation of permeable paths to release water vapor towards the ceramic lining's surfaces during heating.

This review presented and discussed the benefits and drawbacks associated with the use of organic fibers, metallic powders, a special permeability enhancing active compound, silica-based additives and a chelating agent in castable's compositions. The role of each one of them consists of engineering the resultant microstructure and induce the creation of suitable paths to expel the steam generated during the initial stages of the drying process. Although the polymeric fibers are still the most versatile and easy solution commonly used by refractory producers to increase permeability, other interesting and more effective alternative additives have been developed in recent years for this purpose (i.e., permeability enhancing active compound and chelating agents).

Besides that, the optimization of the dry out schedule should take into consideration the set-up of suitable heating rates and dwell times at critical temperature ranges, aiming to favor a more effective water withdrawal during the water evaporation period. Knowing the difficulties to design a proper drying schedule for industrial equipment containing various types of refractory materials, some authors [115] suggested three main recommendations to minimize the lining's spalling risks:

(i) To simulate the likelihood of the dense castable's explosion, by designing tests involving in situ mass loss, temperature and internal pore pressure evaluations.

(ii) To control the drying progress and adjust the heat up schedule, by measuring the inner pore pressure at distinct lining depths and locations (to prevent a critical threshold level being reached).

(iii) To develop dewatering simulation models (using properties such as permeability, thermal conductivity, porosity, water content, etc.) to set up optimized heating curves.

Thus, novel interesting engineering opportunities, such as the use of in-situ based experimental techniques (i.e., neutron and X-ray tomography) to obtain more accurate data and the development of numerical models, might help in simulating and predicting the steam pressure developed in refractory systems during their first heating. Consequently, instead of designing conservative drying schedules based on empirical knowledge, the novel optimized heating procedures should be based on technical and scientific information. A critical analysis of the available mathematical models applied for simulating the castables' drying behavior was discussed in a paper just published by some of the authors [116].

## 4. Acknowledgments

The authors would like to thank Conselho Nacional de Desenvolvimento Cientifico e Tecnologico – CNPq (Finance Code 001 and grant numbers: 303324/2019-8 and 305877/2017-8), Fundação de Amparo à Pesquisa do Estado de São Paulo (FAPESP – grant numbers: 2019/07996-00 and 2018/19773-3) and FIRE (Federation for International Refractory Research and Education) for supporting this work.